\documentclass[aoas,MSNbibl,nameyear,rotating,seceqn,dvips]{arximspdf}
\usepackage{dcolumn}
\makeatletter
\usepackage{mrbibl}
\def\printorigbibl{%
    \def\orig@filename{\jobname.getmref.bbl}
    \IfFileExists{\orig@filename}{

      \pagestyle{empty}
      \csname mrbibl@starthook\endcsname
      \bgroup
        \ifnum\c@page<1000\setcounter{page}{10000}\fi
        \let\bibname\mrbibl@bibname
        \def\bibitem{\let\bibitem\real@bibitem\real@bibitem}
        \let\MR\bbl@MR
        \hideuncited
        \let\PREHOOK@@biblabel\relax
        \let\fmt@empty@citeanchor\@gobble
        \let\fmt@citeanchor\@firstofone
        \@fileswfalse
        \input{\orig@filename}
      \egroup
      \csname mrbibl@endhook\endcsname
      }{}
}
\makeatother

\doi{10.1214/10-AOAS359}
\volume{4}
\issue{2}
\pubyear{2010}
\firstpage{687}
\lastpage{714}

\makeatletter

\newcolumntype{d}[1]{D{.}{.}{#1}}
\newcommand{\eqref}[1]{(\ref{#1})}
\renewcommand{\citep}[1]{(\citeauthor{#1} \citeyear{#1})}
\usepackage{graphicx}
\newcommand{\mixer}{\texttt{MixeR} }
\newcommand{\Esp}{\mathbb{E}}
\newcommand{\pibf}{\bolds{\pi}}
\newcommand{\alphabf}{\bolds{\alpha}}
\newcommand{\betabf}{\bolds{\beta}}
\newcommand{\etabf}{\bolds{\eta}}
\newcommand{\taubf}{\bolds{\tau}}
\newcommand{\Zbf}{\mathbf{Z}}
\newcommand{\Xbf}{\mathbf{X}}
\newcommand{\J}{\mathcal{J}(\Xbf,\mathcal{R}(\Zbf); \betabf)}

\makeatother

\begin{document}
\begin{frontmatter}

\title{Strategies for online inference of model-based clustering in
large and growing networks\protect\thanksref{T1}}
\runtitle{Online methods for model-based clustering on networks}
\thankstext{T1}{Supported in part by French Agence Nationale de la
Recherche Grant NeMo ANR-08-BLAN-0304-01.}

\begin{aug}
\author[a]{\fnms{Hugo} \snm{Zanghi}\ead[label=e1]{hugo.zanghi@exalead.com}},
\author[b]{\fnms{Franck} \snm{Picard}\ead[label=e2]{picard@biomserv.univ-lyon1.fr}\corref{}},
\author[b]{\fnms{Vincent} \snm{Miele}\ead[label=e3]{miele@biomserv.univ-lyon1.fr}}\\
\and
\author[c]{\fnms{Christophe} \snm{Ambroise}\ead[label=e4]{cambroise@genopole.cnrs.fr}}
\runauthor{Zanghi, Picard, Miele and Ambroise}
\affiliation{Laboratoire Statistique et G\'enome and
Exalead, Laboratoire de Biom\'etrie et Biologie Evolutive and
Universite de Lyon, Laboratoire Statistique et G\'enome and
Universite de Lyon and Laboratoire Statistique et G\'enome}
\address[a]{H. Zanghi\\
Exalead\\
10 place de la Madeleine\\
75008 Paris\\ France \\ \printead{e1}}
\address[b]{F. Picard\\
V. Miele\\
Laboratoire Biometrie et Biologie Evolutive\\
UCB Lyon 1---Bat. Gregor Mendel \\
43 bd du 11 novembre 1918\\
69622 Villeurbanne Cedex\\
France\\
E-mails: \printead*{e2}\\
\phantom{E-mails: }\printead*{e3}}

\address[c]{C. Ambroise\\
Laboratoire Statistique et G\'enome\\
UMR CNRS 8071-INRA 1152-UEVE \\
523, place des Terrasses\\
F-91000 Evry \\
France \\
\printead{e4}}
\end{aug}

\received{\smonth{12} \syear{2008}}
\revised{\smonth{4} \syear{2010}}

%
\begin{abstract}
In this paper we adapt online estimation strategies to perform
model-based clustering on large networks. Our work focuses on two
algorithms, the first based on the
SAEM algorithm, and the second on variational methods. These two
strategies are compared with existing approaches on simulated and real data.
We use the method to decipher the connexion structure of the political
websphere during the US political campaign in 2008. We show that our
online EM-based algorithms
offer a good trade-off between precision and speed, when estimating
parameters for mixture distributions in the context of random
graphs.
\end{abstract}

%

\end{frontmatter}

\section{Introduction}

Analyzing networks has become an essential part of a number of
scientific fields. Examples include such widely differing phenomena as power
grids, protein-protein interaction networks and friendship. In this
work we
focus on particular networks which are made of political Weblogs. With
the impact
of new social network websites like Myspace and Facebook, the web has
an increasing
influence on the political debate. As an example, \citet
{adamic2005pba} showed that
blogging played an important role in the political debate of the 2004
US Presidential
Election. Although only a small minority of Americans actually used
these Weblogs,
their influence extended far beyond their readership, as a result of
their interactions
with national mainstream media. In this article we propose to uncover
the connexion
structure of the political websphere during the US political
campaign in 2008.
This data set consists of a one-day snapshot of over 130,520 links and
1870 manually classified
websites (676 liberal, 1026 conservative and 168 independent) where
nodes are connected if there
exists a citation from one to another.

Many strategies have been developed to study networks structure and
topology. A distinction can be made between model-free
[\citet{Newman2006}; \citet{Ng2002}] and model-based methods, with connexions
between parametric and nonparametric models [\citet{BC09}]. Among
model-based methods, model-based clustering has provided
an efficient way to summarize complex networks structures. The basic
idea of these strategies is to model the distribution of connections in
the network,
considering that nodes are spread among an unknown number of
connectivity classes which are
themselves unknown. This generalizes model-based clustering to network
data, and various modeling
strategies have been considered. \citet{nowicki01estimation} propose a
mixture model on dyads that belong
to some relational alphabet, \citet{Daudin2008} propose a mixture on
edges, \citet{Handcock2007} consider continuous
hidden variables and Airoldi et al. (\citeyear{ABX05,ABF07,ABF08}) consider both mixed
membership and stochastic block structure.

In this article our concern is not to assess nor to compare the
appropriateness of these different models, but we focus on a computational
issue that is shared by most of them. Indeed, even if the modeling
strategies are diverse, EM like algorithms constitute a common core of the
estimation strategy [\citet{Dempster1977}; \citet{Snijders1997}], and this
algorithm is known to be slow to convergence and to be very sensitive
to the size of the data set. This issue should be put into perspective
with a new challenge that is inherent to the analysis of network data
sets which is the
development of optimization strategies with a reasonable speed of
execution, and which can deal with networks composed of tens of
thousands of nodes, if
not more. To this extent, Bayesian strategies are limited, as they may
not handle networks with more than a few hundred [\citet
{Snijders1997}; \citet{nowicki01estimation}] or a few thousand [\citet{ABF08}],
and heuristic-based algorithms may not be satisfactory from the
statistical point of view [\citet{Newman2007}]. Variational strategies
have been proposed as well [\citet{ABX05}; \citet{Daudin2008}],
but they are concerned by the same limitations as EM. Thus, the new
question we assess in this work is ``how to perform efficient
model-based clustering
from a computational point of view on very large networks or on
networks that grow over time?''

Online algorithms constitute an efficient alternative to classical
batch algorithms
when the data set grows over time. The application of such strategies
to mixture models
has been studied by many authors [\citet
{Titterington1984}; \citet{WangZhao2006}]. Typical clustering algorithms
include the online $k$-means algorithm [\citet{MacQueen67}]. More
recently, \citet{Liu2006}
modeled Internet traffic using a recursive EM algorithm for the
estimation of Poisson mixture
models. However, an additional difficulty of mixture models for random
graphs is that the computation of
$\Pr\{\Zbf|\Xbf\}$, the distribution of the hidden label variables
$\Zbf$ conditionally
on the observation $\Xbf$, cannot be factorized due to conditional dependency
[\citet{Daudin2008}]. In this work we consider two alternative
strategies to deal
with this issue. The first one is based on the Monte Carlo simulation
of $\Pr\{\Zbf|\Xbf\}$,
leading to a Stochastic version of the EM algorithm (Stochastic
Approximation EM, SAEM)
[\citet{Delyon1999}]. The second one is the variational method proposed
by \citet{Daudin2008} which consists
in a mean-field approximation of $\Pr\{\Zbf|\Xbf\}$. This strategy has
also been proposed by \citet{LATO08X}
and by \citet{ABF08} in the Bayesian framework.

In this article we begin by describing the blog database from the 2008
US presidential campaign.
Then we present the MixNet model proposed by \citet{Daudin2008}, and we
compare the model with its principal
competitors in terms of modeling strategies. We use the \citet
{Sampson68} data set for illustration. We derive the \textit{online}
framework to estimate the parameters of this mixture using
SAEM or variational methods. Simulations are used to show that online
methods are very effective in terms of computation time,
parameter estimation and clustering efficiency. These simulations
integrate both fixed-size and increasing size networks for which
online methods have been designed. Finally, we uncover the
connectivity structure of the 2008 US Presidential websphere using the
proposed variational online algorithm of the MixNet model.

\section{Data presentation}

In this community extraction experiment, we used a data set obtained
on November 7, 2007 by the French company RTGI
(Information Networks, Territories and Geography) using a specific
methodology similar to \citet{fouetillou2007}. This data set consists
of a one-day snapshot of over two thousand websites, one thousand of
which featured in two online directories:
\href{http://wonkosphere.com}{http://wonkosphere.com} and
\href{http://www.politicaltrends.info}{http://www.politicaltrends.info}.\break The first site provides a manual
classification, and the second
an automatic classification based on text analysis. From this seed of
a thousand sites, a web crawler
[\citet{drugeon05}] collected a maximum of 100 pages per hostname which
is in general the sitename. External links were examined to
check the connectivity with visited and unvisited websites. If
websites were still unvisited, and if there existed a minimal path
of distance less than two between a hostname which belongs to the seed
and these websites, then the web crawler collected them.

Using this seed-extension method, 200,000 websites were collected, and
a network of websites was created where nodes represent hostnames
(a hostname contains a set of pages) and edges represent hyperlinks
between different hostnames. Multiple links between two different
hostnames were collapsed into a single link. Intra-domain links were
taken into account if hostnames were not similar.
For this web network, we computed an authority score [\citet
{kleinberg1998}] and a keyword score TF/IDF [\citet{salton1975vsm}] on
focused words
(political entities) in order to identify respectively nodes with
high-quality websites (high authority scores) and centered on those topics
(on a political corpus). 870 new websites emerged out of these two
criteria. They were checked by experts and the validity of the seed confirmed.
The final tally was 130,520 links and 1870 sites: 676 liberal, 1026
conservative and 168 independent. The data can be downloaded at \url
{http://stat.genopole.cnrs.fr/sg/Members/hzanghi}.

\section{A mixture model for networks}


\subsection{Model and notation}

We model the observed network of websites by a random graph $G$, where
$\mathcal{V}$ denotes the set of $n$
fixed vertices which represent hyperlinks between blogs. These random
edges are modeled by $\Xbf=\{ X_{ij}, (i,j)
\in\mathcal{V}^2\}$, a set of random variables coding for the nature
of connection between blogs~$i$ and $j$.
The nature of the links can be discrete or continuous, and we consider
a model with distributions belonging to
the exponential family. In the MixNet model we suppose that nodes are
spread among $Q$ hidden classes and we denote
by $Z_{iq}$ the indicator variable such that $\{Z_{iq}=1\}$ if blog
$i$ belongs to class $q$. We denote by
$\Zbf=(\Zbf_1,\ldots,\Zbf_n)$ the vector of random independent label
variables such that
\[
\Zbf_i \sim\mathcal{M}(1,\alphabf=\{\alpha_1,\ldots,\alpha_Q\}),
\]
with $\alphabf$ the vector of proportions for classes. In the
following, formulas are valid for the case of directed and undirected
networks. Self-loops have not been introduced for simplicity of
notation, and have been implemented in the MixNet software.

\subsubsection*{Conditional distribution} MixNet is defined using the
conditional distribution of edges given
the label of the nodes. $X_{ij}$'s are supposed to be conditionally independent:
\[
\Pr\{\Xbf|\Zbf; \etabf\} = \prod_{ij} \prod_{q,l} \Pr\{X_{ij} |
Z_{iq}Z_{jl}=1 ; \eta_{ql}\} ^{Z_{iq}Z_{jl}},
\]
and $\Pr\{X_{ij}|Z_{iq}Z_{jl}=1;\eta_{ql}\}$ is supposed to belong to
the regular exponential family, with natural parameter $\eta_{ql}$:
\[
\log\Pr\{X_{ij} | Z_{iq}Z_{jl}=1 ; \eta_{ql}\} = \eta_{ql}^{t}
h(X_{ij}) - a(\eta_{ql}) + b(X_{ij}),
\]
where $h(X_{ij})$ is the vector of sufficient statistics, $a$ a
normalizing constant and $b$ a given function.
Consequently, the conditional distribution of
the graph is also from the exponential family:
\[
\log\Pr\{\Xbf|\Zbf; \etabf\} = \sum_{ij,ql} Z_{iq}Z_{jl} \eta
_{ql}^{t} h(X_{ij}) -
\sum_{ij,ql} Z_{iq}Z_{jl}a(\eta_{ql})+ \sum_{ij} b(X_{ij}).
\]
Examples of such distributions are provided in the \hyperref[app]{Appendix}.

\subsubsection*{Models comparison}Many strategies have been considered to
construct models for clustering in networks. Variations mainly
concern the nature of the link between nodes and the definition of
nodes' memberships. For instance, the stochastic blockstructure
model [\citet{Snijders1997}; \citet{nowicki01estimation}] considers links that
are dyads $(X_{ij},X_{ji})$, whereas MixNet considers a model
on edges only. Consequently, MixNet implicitly assumes the
independence of $X_{ij}$ and $X_{ji}$ conditionally on the latent
structure. As for the definition of the label variables, the Mixed
Membership Stochastic Blockmodel (MMSB) has been proposed to describe
the interactions
between objects playing multiple roles [\citet{ABF08}]. Consequently,
the hidden variables of their model can stand for more than one group
for one node, whereas MixNet only considers one label per node. \citet
{ABF08} also model the sparsity of the network. This could be done in
the context of
MixNet by introducing a Dirac mass on zero for the conditional
distribution of edges. Differences among approaches also concern the
statistical framework
that defines subsequent optimization strategies. The Bayesian setting
has been a framework chosen by many authors, as it allows the
integration of prior
information and hierarchical structures [\citet{ABF08}]. On the
contrary, our approach does not necessarily rely on stochastic
strategies, meaning that each run provides the
same set of parameters. However, the likelihood of mixture models in
general is multimodal, which is a problem for both approaches. In MCMC
procedures it leads to
potential label switching issues, and the variational EM may converge
to local maxima.

\begin{sidewaysfigure}

\includegraphics{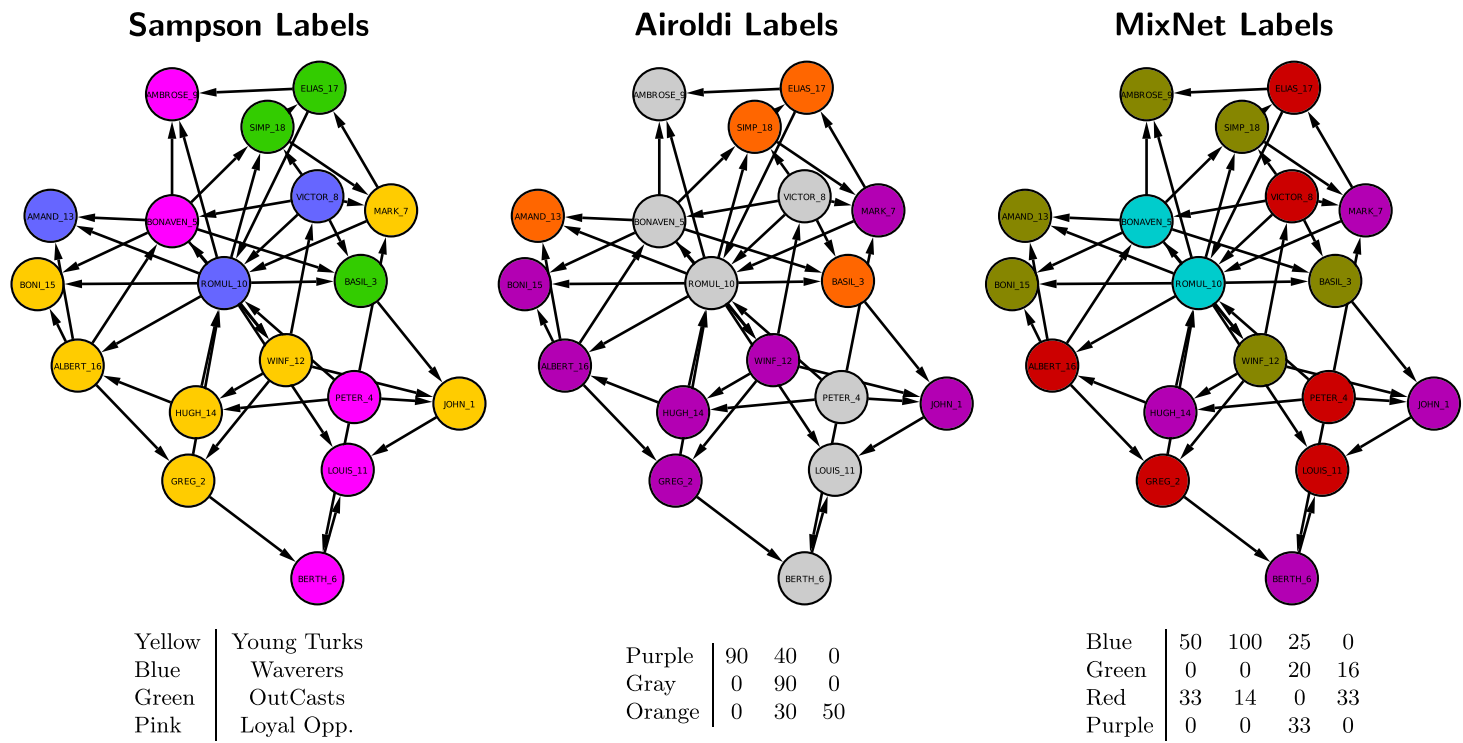}

\caption{Monk data set with different labels: Original categories
obtained by \protect\citet{Sampson68}, Labels obtained by \protect\citet{ABF08},
MixNet labels.
Estimated block model ($\hat{B}$) for MMSB, and estimated
connectivity matrix ($\hat{\pibf}$) for MixNet.}\label{Fig:Sampson}
\end{sidewaysfigure}

As the model and the statistical frameworks are different, clustering
results are likely to be very different as well. In order to
illustrate our point, we deviate from the political blog data and we
use the small data set of
\citet{Sampson68} which is used in \citet{ABF08}. This data set
describes relational data between monks in a monastery (whom do you
like data). Figure \ref{Fig:Sampson}
shows 3 possible partitionings of this graph, the first one
corresponds to Sampson's observations, the second one is the result of
the MMSB model as presented in
\citet{ABF08}, and the third one is provided by MixNet. Individual
labels are provided in Table \ref{Tab:Monk}. As already noted by the
authors, the MMSB classes overlap with the relational categories
provided by Sampson.
This is not the case for MixNet, which uncovers classes of
connectivity that show strong inter-connections but very few
intra-connections ($\hat{\pibf}$).
Since one link exists when a monk likes another, MixNet clusters are
made of monks that like the same sets of other monks. For instance, the
blue cluster
is made of two monks that like each other and that like all monks
assigned to the green cluster. The monks in the green cluster do not
seem to like each other, but
prefer the monks assigned to the red and purple clusters. As a
consequence, both approaches provide different information and are very
complementary with more modeling possibilities in the MMSB
framework, due to the mixed membership and the prior information
integration possibilities. The relevance of MixNet results has been published
elsewhere [\citet{PMD09}], and our aim in this article is not to
compete the models. Our point is rather computational: we aim at
providing an efficient method
to perform model-based clustering on large networks. We use the MixNet
model as a basis for development, but the online framework we develop
could be applied to
the MMSB model as well.

\begin{table}
\caption{Clustering results on the Monk data set. LO---Loyal opponents;
YT---young turks; O---Outcasts; W---waverers}\label{Tab:Monk}
\begin{tabular*}{\textwidth}{@{\extracolsep{\fill}}lccc@{}}
\hline
\textbf{Monk} & \textbf{Sampson label} & \textbf{MMSB label} & \textbf{MixNet label} \\
\hline
Ambrose & LO & Gray & Green\\
Boniface & YT & Violet & Green\\
Mark & YT & Violet & Purple\\
Winfrid & YT & Violet & Green\\
Elias & O & Orange & Red\\
Basil & O & Orange & Green\\
Simplicius & O & Orange & Green\\
Berthold & LO & Gray & Purple\\
John & YT & Violet & Purple\\
Victor & W & Gray & Red\\
Bonaventure & LO & Gray & Blue\\
Amand & W & Orange & Green\\
Louis & LO & Gray & Red\\
Albert & YT & Violet & Red\\
Ramuald & W & Gray & Blue\\
Peter & LO & Gray & Red\\
Gregory & YT & Violet & Red\\
Hugh & YT & Violet & Purple\\
\hline
\end{tabular*}
\end{table}

\subsubsection*{Joint distribution} Since MixNet is defined by its
conditional distribution, we first check that the joint
distribution also belongs to the exponential family. Using notation
\[
\cases{
N_q(\Zbf)  = \displaystyle\sum_i Z_{iq},\vspace*{2pt}\cr
H_{ql}(\Xbf,\Zbf)  =\displaystyle\sum_{ij}Z_{iq}Z_{jl}h(X_{ij}), \vspace*{2pt}\cr
G_{ql}(\Zbf)  =\displaystyle\sum_{ij}Z_{iq}Z_{jl} = N_q(\Zbf) N_l(\Zbf),\vspace*{2pt}\cr
\alpha_q  = \exp(\omega_q) / \displaystyle\sum_l \exp(\omega_l)}
\]
and
\[
\cases{
T(\Xbf,\Zbf)  = ( \{N_q(\Zbf)\}, \{H_{ql}(\Xbf,\Zbf)\}, \{
G_{ql}(\Zbf)\}), \vspace*{2pt}\cr
\betabf  = ( \{\omega_q\}, \{\eta_{ql}\}, \{-a(\eta_{ql})\}
), \vspace*{2pt}\cr
A(\betabf)  = n \log\displaystyle\sum_l \exp\omega_l, \vspace*{2pt}\cr
B(\Xbf)  = \displaystyle\sum_{ij} b(X_{ij}),
}
\]
we have the factorization $\log\Pr\{\Xbf,\Zbf; \betabf\} =
\betabf^{t} T(\Xbf,\Zbf) - A(\betabf)+B(\Xbf),
$ which proves the claim. The sufficient statistics $T(\Xbf,\Zbf)$ of
the complete-data model are the number
of nodes in the classes $N_q(\Zbf)$, the characteristics of the
between-group links ($H_{ql}$ through function $h$ that can stand for
the number of between group links or for the intensity of the
connections in the case of edges
with Poisson or Gaussian distributions), and the product of
frequencies between classes $G_{ql}$. In the following we aim at
estimating $\betabf$.

\subsection{Sufficient statistics and online recursion}

Online algorithms are incremental algorithms which recursively update
parameters, using current
parameters and new observations. We introduce the following notation.
Let us denote by $\Xbf^{[n]} = \{X_{ij}\}_{i,j=1}^{n}$
the adjacency matrix of the data, when $n$ nodes are present, and by
$\Zbf^{[n]}$ the associated labels.
A convenient notation in this context is $\Xbf_{i,\bullet}= \{X_{ij},
 j \in\mathcal{V}\}$, which denotes all the edges related to node
$i$. Note that the addition of one node leads to the addition of $n+1$
potential connections.

The use of online methods is based on the additivity of the sufficient
statistics regarding the addition of
a new node. We can show that
\[
\cases{
N_q\bigl(\Zbf^{[n+1]}\bigr)  = N_q\bigl(\Zbf^{[n]}\bigr) + Z_{n+1,q},\vspace*{2pt}\cr
H_{ql}\bigl(\Xbf^{[n+1]},\Zbf^{[n+1]}\bigr)  = H_{ql}\bigl(\Xbf^{[n]},\Zbf^{[n]}\bigr) +
\xi_{ql}^{[n+1]}, \vspace*{2pt}\cr
G_{ql}\bigl(\Zbf^{[n+1]}\bigr)  = G_{ql}\bigl(\Zbf^{[n]}\bigr) + \zeta_{ql}^{[n+1]},
}
\]
with
\begin{eqnarray*}
\xi_{ql}^{[n+1]} & = & Z_{n+1,q} \sum_{j=1}^n Z_{jl}h(X_{n+1,j}) +
Z_{n+1,l} \sum_{i=1}^n Z_{iq}h(X_{i,n+1}), \\
\zeta_{ql}^{[n+1]} & = & Z_{n+1,q} N_{l}^{[n]} + Z_{n+1,l} N_{q}^{[n]}
+ Z_{n+1,q}I\{q=l\}.
\end{eqnarray*}
Then if we define $T(\Xbf_{n+1,\bullet},\Zbf^{[n+1]})=(Z_{n+1,q},\{
\xi_{ql}^{[n+1]}\},\{\zeta_{ql}^{[n+1]}\})$, we get
%
\begin{equation}
\label{Eq:addit}
T\bigl(\Xbf^{[n+1]},\Zbf^{[n+1]}\bigr) = T\bigl(\Xbf^{[n]},\Zbf^{[n]}\bigr) + T\bigl(\Xbf
_{n+1,\bullet},\Zbf^{[n+1]}\bigr).
\end{equation}

Those equations will be used for parameter updates in the online algorithms.

\subsection{Likelihoods and online inference}

Existing estimation strategies are based on maximum likelihood, and
algorithms related to EM are used for
optimization purposes. The aim is to maximize the conditional
expectation of the complete-data log-likelihood
\[
\mathcal{Q}(\betabf|\betabf^{\prime})  =  \sum_{\Zbf} \Pr\{\Zbf|\Xbf
; \betabf'\} \log\Pr\{\Xbf,\Zbf;\betabf\},
\]
\noindent and the main difficulty is that $\Pr\{\Zbf|\Xbf;\betabf'\}$
cannot be factorized and needs to be approximated [\citet{Daudin2008}]. A
first strategy to simplify the problem is to consider a classification
EM-based strategy [\citet{CG92}]. In this setting label variables are
considered as nonrandom and are replaced by their prediction (0$/$1).
This is a generalization of the $k$-means algorithm for which the
problem of
computing $\Pr\{\Zbf|\Xbf\}$ is left apart. This strategy has been the
subject of a previous work
[\citet{Zanghi2007}]. It is known to give biased estimates, but is very
efficient from a computational time point of view.

To this strategy, we propose two different alternatives based on the
Stochastic Approximation EM approach [\citet{Delyon1999}] which approximates
$\Pr\{\Zbf|\Xbf\}$ using Monte Carlo simulations, and on the so-called
variational approach, which consists of approximating $\Pr\{\Zbf|\Xbf\}$
by a more tractable distribution on the hidden variables. In their
online versions, these algorithms optimize $\mathcal{Q}(\betabf|\betabf
^{\prime})$
sequentially, while nodes are added. To this extent, we introduce notation
\begin{eqnarray*}
\mathcal{Q}_{n+1}\bigl(\betabf|\betabf^{[n]}\bigr) =  \sum_{\Zbf^{[n+1]}} \Pr
\bigl\{\Zbf^{[n+1]}|\Xbf^{[n+1]}; \betabf^{[n]}\bigr\} \log\Pr\bigl\{\Xbf^{[n+1]},\Zbf
^{[n+1]};\betabf\bigr\},
\end{eqnarray*}
with $[n+1]$ being either the number of nodes or the increment of the
algorithm, which are identical in the online context.

\eject
\section{Stochastic approximation EM for network mixture}

\subsection{A short presentation of SAEM}
An original way of estimating the parameters of the MixNet model is to
approximate the expectation of the complete data log-likelihood
using Monte Carlo simulations corresponding to the Stochastic
Approximation EM algorithm [\citet{Delyon1999}]. In situations where maximizing
$\mathcal{Q}(\betabf|\betabf')$ is not in a simple closed form, the
SAEM algorithm maximizes an approximation $\widehat{\mathcal{Q}}(\betabf
|\betabf')$
computed using standard stochastic approximation theory such that
%
\begin{equation}
\label{eq:saem}
\widehat{\mathcal{Q}}(\betabf|\betabf')^{[k]} = \widehat{\mathcal
{Q}}(\betabf|\betabf')^{[k-1]} +
\rho_k \bigl(\widetilde{\mathcal{Q}}(\betabf|\betabf') - \widehat
{\mathcal{Q}}(\betabf|\betabf')^{[k-1]}\bigr),
\end{equation}
where $k$ is an iteration index, $\{\rho_k\}_{k \geq1}$ a sequence of
positive step size and where $\widetilde{Q}(\betabf|\betabf')$ is
obtained by Monte
Carlo integration. This is a simulation of the expectation of the
complete log-likelihood
using the posterior $\Pr\{\Zbf|\Xbf\}$. Each iteration $k$ of the
algorithm is broken down into three steps:
\begin{description}
\item[Simulation] of the missing data. This can be achieved using
Gibbs Sampling of the posterior $\Pr\{\Zbf|\Xbf\}$.
The result at iteration number $k$ is $m(k)$ realizations of
the latent class data $\Zbf$: $(\mathbf{Z}({1}),\ldots,
\mathbf{Z}({m(k)}))$.
\item[Stochastic approximation] of $\mathcal{Q}(\betabf|\betabf')$
using equation (\ref{eq:saem}), with
%
\begin{equation}
\widetilde{\mathcal{Q}}(\betabf|\betabf') =
\frac{1}{m(k)} \sum_{s=1}^{m(k)} \log\Pr(\mathbf{X},\mathbf
{Z}(s);\betabf).
\end{equation}
\item[Maximization] of $\widehat{\mathcal{Q}}(\betabf|\betabf')^{[k]}$
according to $\betabf$.
\end{description}

As regards the online version of the algorithm, the number of
iterations $k$ usually coincides with $n+1$, the number of nodes of the
network. Although it is possible to go further in the iterative
process to improve the estimates, it is rarely necessary since the
results obtained with $n+1$ iterations are usually reliable. This can
be explained by the fact that the MixNet model is robust to sampling.
The information in the network is indeed highly redundant and a
reliable estimation of the network parameters can be obtained with a
small sample (a few dozen) of the nodes using a classical batch
algorithm. When $n$ is large, using an online algorithm with all the
nodes is similar to performing many iterations of a batch algorithm on
a small sample.


\subsection{Simulation of $\Pr\{\Zbf|\Xbf\}$ in the online
context}\label{sec:simulation}
We use Gibbs sampling which is applicable when the joint distribution
is not known explicitly,
but the conditional distribution of each variable is known. Here we
generate a sequence of~$\Zbf$ approaching
$\Pr\{\Zbf|\Xbf\}$ using $\Pr\{Z_{iq}=1 |\mathbf{X}, \mathbf
{Z}_{\backslash i}\}$, where
$\mathbf{Z}_{\backslash i}$ stands for the class of all nodes except
node $i$. The sequence of
samples is a Markov chain, and the stationary distribution of this
Markov chain corresponds
precisely to the joint distribution we wish to obtain. In the online
context, we consider only one simulation to simulate the
class of the last incoming node using
\begin{eqnarray*}
&&\Pr\bigl\{Z_{n+1,q}=1 | \mathbf{X}^{[n+1]},\Zbf^{[n]}\bigr\}\\
&&\qquad = \frac{\Pr\{
Z_{n+1,q}=1,\Zbf^{[n]},\mathbf{X}^{[n+1]}\}}{\sum_{\ell=1}^Q \Pr\{
Z_{n+1,\ell}=1,\Zbf^{[n]},\mathbf{X}^{[n+1]}\} }. \\
&&\qquad =  \frac{\exp\{\betabf^t T(\Xbf_{n+1,\bullet},\Zbf
^{[n]},Z_{n+1,q}) \}}{\sum_{\ell=1}^Q\exp\{\betabf^t T(\Xbf
_{n+1,\bullet},\Zbf^{[n]},Z_{n+1,\ell})\}} \\
&&\qquad \propto \exp\Biggl( \omega_q + \sum_{\ell=1}^Q \eta_{q \ell} \sum
_{j=1}^n Z_{j \ell}h(X_{n+1,j}) + \sum_{\ell=1}^Q N_{\ell}\bigl(\Zbf^{[n]}\bigr)
a(\eta_{q\ell})\Biggr).
\end{eqnarray*}


\subsection{Computing $\widehat{\mathcal{Q}}(\mathbf{\beta} | \mathbf
{\beta}' )$ in the online context}
As regards the online version of the SAEM algorithm, the difference
between the old and the new complete-data log-likelihood may be
expressed as
\begin{eqnarray*}
&&\log\Pr\bigl(\Xbf^{[n+1]},\Zbf^{[n+1]},\betabf\bigr) - \log\Pr\bigl(\Xbf^{[n]},\Zbf
^{[n]},\betabf\bigr) \\
&&\qquad = \log\alpha_q + \sum_{l, i<n+1} Z_{il} \log\Pr\{ X_{n+1,i}
| Z_{n+1,q} Z_{il} \},
\end{eqnarray*}
where the added simulated vertex label is equal to $q$
($Z_{n+1,q}=1$).

Recall that in the online framework, the label of the new node has
been sampled from the Gibbs sampler described in Section \ref
{sec:simulation}. Consequently,
only one possible label is considered in this equation. Then a natural
way to adapt equation (\ref{eq:saem}) to the online context is to approximate
\[
\widetilde{\mathcal{Q}}_{n+1}\bigl(\betabf|\betabf^{[n]}\bigr) -
\widehat{\mathcal{Q}}_{n}\bigl(\betabf|\betabf^{[n]}\bigr)
\]
by
\[
\log\Pr\bigl(\Xbf^{[n+1]},\Zbf^{[n+1]},\betabf\bigr) - \log\Pr\bigl(\Xbf^{[n]},\Zbf
^{[n]},\betabf\bigr).
\]
Indeed, this quantity corresponds to the difference between the
log-likelihood of the original network and log-likelihood of the new
network including the additional node. Notice that the larger the
network, the larger its
associated complete expected log-likelikelihood. Thus, $\log\Pr(\Xbf
^{[n+1]},\Zbf^{[n+1]},\betabf)$ becomes smaller and smaller compared to
${Q}(\betabf| \betabf')$ as $n$ increases. The decreasing
step $\rho_n$ is thus set to one in this online context. We propose
the following update equation for stochastic online EM computation of
the MixNet conditional expectation:
\[
\widehat{\mathcal{Q}}_{n+1}\bigl(\betabf| \betabf^{[n]}\bigr) =
\widehat{\mathcal{Q}}_{n}\bigl(\betabf|\betabf^{[n]}\bigr) +
\log\alpha_q + \sum_{l,i<n+1} Z_{il} \log\Pr\{ X_{n+1,i}
| Z_{n+1,q} Z_{il} \},
\]
where $\Zbf_{n+1}$ is drawn from the Gibbs sampler.

\subsection{Maximizing $\widehat{\mathcal{Q}}(\mathbf{\beta} | \mathbf
{\beta}' )$, and parameters update}

The principle of online algorithms is to modify the current parameter
estimation using the information added by a new available $[n+1]$ node
and its corresponding connections $\Xbf_{n+1,\bullet}$ to the already
existing network. Maximizing $\widehat{\mathcal{Q}}_{n+1}(\betabf|
\betabf^{[n]})$ according to $\betabf$ is straightforward and produces
the maximum likelihood estimates for iteration $[n+1]$. Here we have
proposed a simple version of the algorithm by setting the number of
simulations to one ($m(k)=1$). In this context, the difference between
$\widehat{\mathcal{Q}}_{n}(\betabf|\betabf^{[n]})$ and
$\widehat{\mathcal{Q}}_{n+1}(\betabf| \betabf^{[n]})$ implies only
the terms of the complete log-likelihood which are a function of node
$n+1$. Using notation $ \psi_{ql} = \frac{\partial
a(\eta_{ql})}{\partial\eta_{ql}}, $ we get
\[
\cases{
\alpha_q^{[n+1]}  = N_q\bigl(\Zbf^{[n+1]}\bigr)/(n+1),\vspace*{2pt}\cr
\psi_{ql}^{[n+1]}  = H_{ql}\bigl(\Xbf^{[n+1]},\Zbf^{[n+1]}\bigr)/G_{ql}\bigl(\Zbf^{[n+1]}\bigr),
}
\]
where $(\xi_{ql},\zeta_{ql})$ were defined in the previous section.
Notice that updating the function $\psi_{ql}$ of the parameter of
interest is often more convenient in an online context than directly
considering this parameter of interest. An example of parameter
update is given for the Bernoulli and Poisson cases in the \hyperref[app]{Appendix}.

Once all the nodes in the network have been visited (or are known),
the parameters can be further improved and the complete
log-likelihood better approximated by continuing with the SAEM
algorithm described above.


\section{Application of \textit{online} algorithm to variational methods}

Variational methods constitute an alternative to SAEM. Their principle
is to approximate
the untractable distribution $\Pr\{\Zbf| \Xbf;\betabf\}$ by a newly introduced
distribution on $\Zbf$ denoted by $\mathcal{R}$. Then this new
distribution is used to optimize
$\J$, an approximation (lower bound) of the incomplete-data
log-likelihood $\log\Pr\{ \Xbf;\betabf\}$, defined such that
\[
\J  =  \log\Pr\{\Xbf; \betabf\} - \mathit{KL}(\mathcal{R}(\Zbf), \Pr\{
\Zbf|\Xbf; \betabf\}),
\]
with $\mathit{KL}(\bullet|\bullet)$ being the Kullback--Leibler divergence
between probability distributions [\citet{Jordan1999}].
Then one must choose the form of $\mathcal{R}$, and the product of
Multinomial distributions is natural in the case of MixNet, with $\log
\mathcal{R}(\Zbf) = \sum_i \sum_q Z_{iq} \log\tau_{iq},$
and the constraint $\sum_q \tau_{iq}=1$. In this case, the form of $\J
$ is
\begin{eqnarray*}
\J& = & \sum_{\Zbf} \mathcal{R}(\Zbf; \taubf) \log\Pr\{\Xbf,\Zbf
;\betabf\} - \sum_{\Zbf} \mathcal{R}(\Zbf; \taubf) \log \mathcal
{R}(\Zbf; \taubf)\\
& = & \mathcal{Q}(\taubf, \betabf) + \mathcal{H}(\mathcal{R}(\Zbf;
\taubf)),
\end{eqnarray*}
with $\mathcal{Q}(\taubf, \betabf)$ an approximation of the conditional
expectation of the complete-data
log-likelihood, and $\mathcal{H}(\mathcal{R}(\Zbf; \taubf))$ the
entropy of the approximate \textit{posterior}
distribution of $\Zbf$.

The implementation of variational methods in online algorithms relies
on the additivity property of $\J$ when nodes
are added. This property is straightforward: $\mathcal{Q}(\taubf,
\betabf)$ is additive thanks to equation (\ref{Eq:addit})
[because $\mathcal{R}(\Zbf)$ is factorized], and $\mathcal{H}(\mathcal
{R}(\Zbf; \taubf))$ is also additive, since
the hidden variables are supposed independent under $\mathcal{R}$ and
the entropy of independent variables is additive.
The variational algorithm is very similar to an EM algorithm, with the
E-step being replaced by a variational step
which aims at updating variational parameters. Then a standard M-step
follows. In the following, we give the
details of these two steps in the case of a variational online algorithm.

\subsection{Online variational step}
When a new node is added, it is necessary to compute its associated
variational parameters $\{\tau_{n+1,q}\}_q$. If we consider all
the other $\tau_{iq}$ for $i < n+1$ as known, the $\{\tau_{n+1,q}\}_q$
are obtained by differentiating the criterion
\[
\mathcal{J}\bigl( \Xbf^{[n+1]},\mathcal{R}\bigl(\Zbf^{[n+1]}\bigr) ; \betabf
\bigr) + \sum_{i=1}^{n+1} \Lambda_i \Biggl(\sum_{q=1}^Q \tau_{iq}-1 \Biggr),
\]
where the $\Lambda_i$ are the Lagrangian parameters. Since function
$\mathcal{J}$ is additive according to the nodes, the calculation of
its derivative according to
$\tau_{n+1,q}$ gives
\[
\omega_{q}^{[n]} + \sum_{l=1}^{Q}\sum_{j=1}^{n} \tau_{jl}^{[n]} \bigl(
\eta_{ql}^{[n]} h(X_{n+1,j}) + a\bigl(\eta_{ql}^{[n]}\bigr) \bigr) - \log\tau
_{n+1,q} +1 + \Lambda_{n+1} = 0.
\]
This leads to
%
\begin{eqnarray}\label{logit}
&&\tau_{n+1,q} \propto\alpha_q^{[n]} \exp\Biggl\{ \sum_{l=1}^{Q}\sum
_{j=1}^{n} \tau_{jl}^{[n]} \bigl( \eta_{ql}^{[n]} h(X_{n+1,j}) + a\bigl(\eta
_{ql}^{[n]}\bigr) \bigr) \Biggr\}
\nonumber
\\[-8pt]
\\[-8pt]
\eqntext{ \forall q \in\{1,\ldots,Q\}.}
\end{eqnarray}

\subsection{Maximization/update step}
To maximize the approximated expectation of the complete
log-likelihood according to $\betabf$, we solve
%
\begin{equation}
\label{onlinederiv}
\frac{\partial\mathcal{Q}_{n+1}(\taubf, \betabf)}{\partial\betabf} =
\Esp_{\mathcal{R}^{[n]}} \biggl(\frac{\partial\log\Pr\{\Xbf^{[n+1]} ,
\Zbf^{[n+1]} ;\betabf\}}{\partial\betabf}\biggr) = 0.
\end{equation}
Differentiating equation (\ref{onlinederiv}) with respect to parameters
$\{\omega_q\}$ gives the following update equation:
\[
\alpha_q^{[n+1]} = \frac{1}{n+1} \Biggl( \sum_{i=1}^{n}\tau_{iq}^{[n]} +
\tau_{n+1,q} \Biggr).
\]
The other update equation is obtained by considering parameters $\{\eta
_{ql}\}$, and using notation $\psi_{ql}$, which gives
\[
{\psi}_{ql}^{n+1} = \frac{\Esp_{\mathcal{R}^{[n]}}(H_{ql}(\Xbf
^{[n+1]},\Zbf^{[n+1]}))}{\Esp_{\mathcal{R}^{[n]}}(G_{ql}(\Zbf
^{[n+1]}))}.
\]
Thanks to equation (\ref{Eq:addit}), which gives the relationships
between sufficient statistics at two successive iterations, parameters
can be computed
recursively using the update of the expectation of the sufficient
statistics, such that
\begin{eqnarray*}
\label{eq:3}
\Esp_{\mathcal{R}^{[n]}} \bigl(N_q\bigl(\Zbf^{[n+1]}\bigr) \bigr) & = & \Esp
_{\mathcal{R}^{[n]}} \bigl( N_q\bigl(\Zbf^{[n]}\bigr) \bigr)+ \Esp_{\mathcal
{R}^{[n]}} ( Z_{n+1,q} ),\\
\Esp_{\mathcal{R}^{[n]}} \bigl( H_{ql}\bigl(\Xbf^{[n+1]},\Zbf^{[n+1]}\bigr)
\bigr) & = & \Esp_{\mathcal{R}^{[n]}} \bigl( H_{ql}\bigl(\Xbf^{[n]},\Zbf^{[n]}\bigr)
\bigr)+ \Esp_{\mathcal{R}^{[n]}} \bigl( \xi_{ql}^{[n+1]} \bigr),\\
\Esp_{\mathcal{R}^{[n]}} \bigl( G_{ql}\bigl(\Zbf^{[n+1]}\bigr)\bigr) & = & \Esp
_{\mathcal{R}^{[n]}} \bigl( G_{ql}\bigl(\Zbf^{[n]}\bigr) \bigr)+ \Esp_{\mathcal
{R}^{[n]}} \bigl( \zeta_{ql}^{[n+1]}\bigr).
\end{eqnarray*}
An example of parameters update is given in the \hyperref[app]{Appendix} for both the
Bernoulli and the Poisson distributions. Note the similarity of the
formula compared with the SAEM strategy. Hidden variables $\Zbf$ are
either simulated
or replaced by their approximated conditional expectation (variational
parameters).

\section{Experiments}

\subsubsection*{Motivations} Experiments are carried out to assess the
trade-off established by online algorithms in terms of
quality of estimation and speed of execution. We propose a
two-online-step simulation study. We first report simulation
experiments using synthetic data generated according to the assumed
random graph model. In this first experiment
we use a simple affiliation model to check precisely the quality of
the estimations given by the online algorithms.
Results are compared to the batch variational EM proposed by \citet
{Daudin2008} to assess the effect of the online
framework on the estimation quality and on the speed of execution. In
a second step, we use a real data set from the
web as a starting point to simulate growing networks with complex
structure, and to assess the performance of online
methods on this type of network. An ANSI \texttt{C}$++$ implementation of
the algorithms is available at
\href{http://stat.genopole.cnrs.fr/software/mixnet/}{http://stat.genopole.cnrs.fr/software/mixnet/}, as well as an R
package named \mixer(\url
{http://cran.r-project.org/web/packages/mixer/}), along with public
data sets.
This software is currently used by the \textit{Constellations} online
application (\url{http://constellations.labs.exalead.com/}),
which instantaneously extracts, visually explores and takes advantages
of the MixNet algorithm to reveal the connectivity
information induced by hyperlinks between the first hits of a given
search request.

\subsection{Comparison of algorithms}

\subsubsection*{Simulations set-up} 
\begin{figure}

\includegraphics{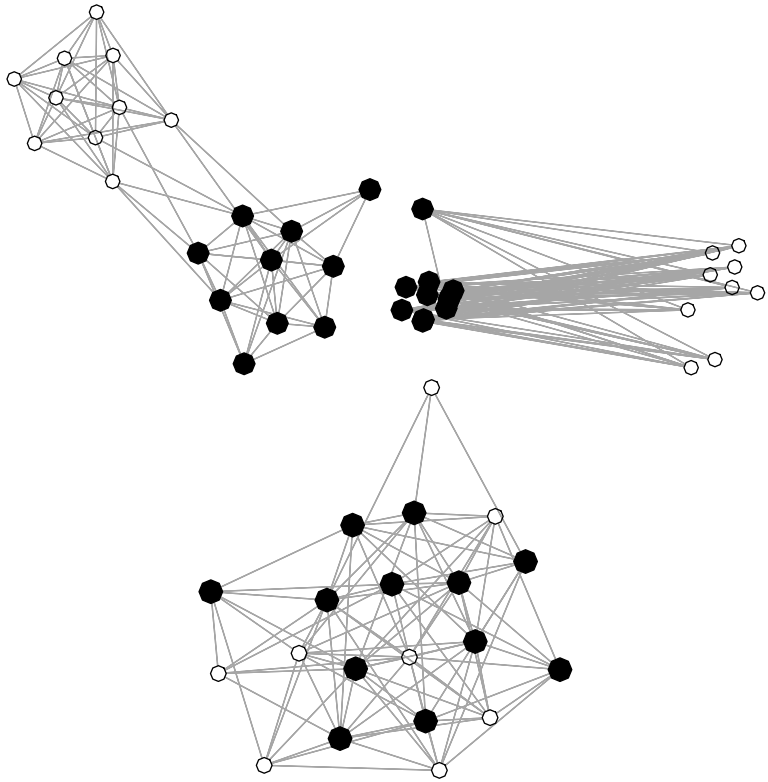}

\caption{Top left: low inter-module
connectivity and strong intra-module connectivity (model 1).
Top right: strong inter-module connectivity and low intra-module
connectivity (model 5). Bottom center:
Erd\H{o}s--R\'enyi model (model 4).}\label{figure:graphs}
\end{figure}

We simulate affiliation models with
$\lambda$ and $\varepsilon$ being the within and between
group probability of connection respectively. Five models are
considered (Table \ref{table:models}). We set $\lambda=1-\varepsilon$
to reduce the number of free parameters, with parameter $\lambda$
controlling the complexity of the model. Differences between
models lie in their modular structure which varies from no structure
(almost the Erd\H{o}s--R\'enyi model) to strong modular structure
(low inter-module connectivity and strong intra-module connectivity,
or strong inter-module connectivity and low intra-module connectivity).
Figure \ref{figure:graphs} illustrates three
kinds of connectivity which allows to represent graphically model 1, 4 and 5.
For each affiliation model we generate graphs with $Q \in\{
2,5,20\}$ groups mixed in the same proportions $1/Q$.
The number of nodes $n$ varies in $\{
100,250,500,750,1000,2000\}$ to explore different sizes of
graphs. We generate
a total of 45 graph models, each being simulated 30 times.

%
\begin{table}[b]
\caption{Parameters of the five affiliation
models considered in the experimental setting} \label{table:models}
\begin{tabular*}{100pt}{@{\extracolsep{\fill}}lcc@{}}
\hline
\textbf{Model} & \multicolumn{1}{c}{$\bolds{\varepsilon}$} & \multicolumn{1}{c@{}}{$\bolds{\lambda}$}\\
\hline
1 & 0.3\phantom{0} & 0.7\phantom{0} \\
2 & 0.35 & 0.65 \\
3 & 0.4\phantom{0} & 0.6\phantom{0}\\
4 & 0.5\phantom{0} & 0.5\phantom{0} \\
5 & 0.9\phantom{0} & 0.1\phantom{0} \\
\hline
\end{tabular*}
\end{table}


\subsubsection*{Criteria of comparison} The comparison between algorithms
is done using the bias $\mathbb{E}(\varepsilon-\widehat{\varepsilon})/\varepsilon
$ and
the mean square error $\mathbb{V}(\widehat{\varepsilon})$ to reflect
estimators variability. We also use the adjusted Rand Index [\citet
{huar85}] to
evaluate the agreement between the estimated and the actual
partitions. Computing this index is based on a ratio between the number
of node
pairs belonging to the same and to different classes when considering
the actual partition and the estimated partition. It lies between 0 and 1,
two identical partitions having an adjusted Rand Index equal to 1.

\subsubsection*{Algorithms set-up} In a first step we compete algorithms
that are based on maximum likelihood estimation (MLE).
The online SAEM and online variational method we propose are compared
with the variational method proposed in \citet{Daudin2008}
(batch MixNet in the sequel). We also add an online classification
version (online CEM) in the comparison since this strategy
has been shown to reduce the computational cost as well [\citet
{Zanghi2007}]. To avoid initialization issues, each algorithm is
started with the same strategy: multiple initialization points are
proposed and the best result is selected based on its likelihood.
The number of clusters is chosen using the Integrated Classification
Likelihood criterion, as proposed in \citet{Daudin2008}.
The algorithms are stopped when the parameters are stable between two
consecutive iterations. In a second step, we compare the
MLE-based algorithms with other competitors like spectral clustering
[\citet{Ng2002}] and a $k$-means like algorithm [\citet{Newman2006}].

\subsubsection*{Estimators bias and MSE (Table \textup{\protect\ref{table:analyseE}})} A
first result is that every algorithm provides estimators with
negligible bias (lower than $1\%$)
and variance for highly structured models (models 1, 2, 5, Table \ref
{table:analyseE}). The online framework shows its limitations when
the structure of the network is less pronounced (model 3), as every
online method shows a significant bias and low precision, whereas the
batch MixNet behaves well. This limitation was expected, as the gain
in computational burden has an impact on the complexity of structures
that can be identified. Finally, among online versions of the
algorithm, the online variational method provides the best results on average
in terms of bias and precision.
%
\begin{table}
\tabcolsep=0pt
\caption{Bias (in percent) and Root Mean Square
Errors ($\times\!10^3$) for the parameters estimators in the five
affiliation models. The $Q$ modules are mixed in the same proportion.
Each model considers $n=500$ nodes and $Q=5$ groups}\label{table:analyseE}
\begin{tabular*}{\textwidth}{@{\extracolsep{\fill}}ld{2.2}d{3.2}d{2.2}d{3.2}d{2.2}d{3.2}d{2.2}d{2.2}@{}}
\hline
& \multicolumn{2}{c}{\textbf{Online-SAEM}}& \multicolumn
{2}{c}{\textbf{Online-variational}}& \multicolumn
{2}{c}{\textbf{Online-CEM}}&\multicolumn{2}{c@{}}{\textbf{Batch-MixNet}}\\[-6pt]
& \multicolumn{2}{c}{\hrulefill}& \multicolumn
{2}{c}{\hrulefill}& \multicolumn
{2}{c}{\hrulefill}&\multicolumn{2}{c@{}}{\hrulefill}\\
\textbf{Model} & \multicolumn{1}{c}{$\mathbf{B}_{\bolds{\%}}\!\bolds{(\widehat{\varepsilon})}$} &
\multicolumn{1}{c}{$\mathbf{B}_{\bolds{\%}}\!\bolds{(\widehat
{\lambda})}$} & \multicolumn{1}{c}{$\mathbf{B}_{\bolds{\%}}\!\bolds{(\widehat{\varepsilon})}$} &
\multicolumn{1}{c}{$\mathbf{B}_{\bolds{\%}
}\!\bolds{(\widehat{\lambda})}$}&
\multicolumn{1}{c}{$\mathbf{B}_{\bolds{\%}}\!\bolds{(\widehat{\varepsilon})}$} &
\multicolumn{1}{c}{$\mathbf
{B}_{\bolds{\%}}\!\bolds{(\widehat{\lambda})}$} &
\multicolumn{1}{c}{$\mathbf{B}_{\bolds{\%}}\!\bolds{(\widehat{\varepsilon})}$} &
\multicolumn{1}{c@{}}{$\mathbf{B}_{\bolds{\%}}\!\bolds{(\widehat{\lambda})}$}\\
\hline
1 & -0.14 & 0.04 & -0.13 & 0.04 & -0.13 & 0.04 & -0.13 & 0.04 \\
2 & 0.23 & -1.01 & 0.04 & -0.11 & -0.03 & 0.01 & -0.03 & 0.00 \\
3 & 9.47 & -26.38 & 8.83 & -24.32 & 6.44 & -22.46 & -0.01 & -0.11 \\
4 & 1.11 & -4.29 & 0.16 & -0.35 & 3.00 & -4.32 & 0.05 & -0.01 \\
5 & -0.01 & -0.02 & -0.01 & -0.02 & -0.01 & -0.02 & -0.01 & -0.02 \\[12pt]
& \multicolumn{1}{c}{RMSE($\widehat{\varepsilon})$} &
\multicolumn{1}{c}{RMSE($\widehat{\lambda})$} &
\multicolumn{1}{c}{RMSE($\widehat{\varepsilon})$} &
\multicolumn{1}{c}{RMSE($\widehat{\lambda})$}&
\multicolumn{1}{c}{RMSE($\widehat{\varepsilon})$} &
\multicolumn{1}{c}{RMSE($\widehat{\lambda})$} &
\multicolumn{1}{c}{RMSE($\widehat{\varepsilon})$} &
\multicolumn{1}{c@{}}{RMSE($\widehat{\lambda})$}\\[3pt]
1 & 1.45 & 2.25 & 1.42 & 2.25 & 1.45 & 2.25 & 1.45 & 2.25 \\
2& 1.89 & 4.04 & 1.65 & 2.90 & 1.63 & 2.90 & 1.63 & 2.90 \\
3& 5.19 & 14.75 & 6.95 & 22.32 & 13.89 & 25.96 & 2.14 & 6.74 \\
4& 3.75 & 10.42 & 1.33 & 1.67 & 8.21 & 15.71 & 1.25 & 1.62 \\
5& 0.92 & 1.73 & 0.92 & 1.73 & 0.93 & 1.73 & 0.92 & 1.73 \\
\hline
\end{tabular*}
\end{table}


\begin{table}
\caption{Means and standard deviations of the
Rand Index for all models with $q$ and $n$ fixed}\label{table:analyseM}
\begin{tabular*}{\textwidth}{@{\extracolsep{\fill}}lcccccccr@{}}
\hline
& \multicolumn{2}{c}{\textbf{Online-SAEM}} & \multicolumn{2}{c}{\textbf{Online-variational}} & \multicolumn{2}{c}{\textbf{Online-CEM}} & \multicolumn
{2}{c@{}}{\textbf{Batch-MixNet}} \\[-6pt]
& \multicolumn{2}{c}{\hrulefill} & \multicolumn{2}{c}{\hrulefill} & \multicolumn{2}{c}{\hrulefill} & \multicolumn
{2}{c@{}}{\hrulefill} \\
\textbf{Model} & \multicolumn{1}{c}{$\bolds{\overline{\mathit{rand}}}$} &
\multicolumn{1}{c}{$\bolds{\sigma_{\mathit{rand}}}$} &
\multicolumn{1}{c}{$\bolds{\overline{\mathit{rand}}}$} &
\multicolumn{1}{c}{$\bolds{\sigma_{\mathit{rand}}}$} &
\multicolumn{1}{c}{$\bolds{\overline{\mathit{rand}}}$} &
\multicolumn{1}{c}{$\bolds{\sigma_{\mathit{rand}}}$} &
\multicolumn{1}{c}{$\bolds{\overline{\mathit{rand}}}$} &
\multicolumn{1}{r@{}}{$\bolds{\sigma_{\mathit{rand}}}$} \\
\hline
1 & 0.98 & 0.02 & 0.98 & 0.02 & 0.98 & 0.02 & 0.99 & 0.02\\
2 & 0.96 & 0.07 & 0.97 & 0.07 & 0.97 & 0.07 & 0.98 & 0.01\\
3 & 0.13 & 0.13 & 0.10 & 0.15 & 0.25 & 0.16 & 0.85 & 0.14\\
4 & 0.00 & 0.00 & 0.00 & 0.00 & 0.00 & 0.00 & 0.00 & 0.00\\
5 & 1\phantom{00} & 0.00 & 1\phantom{00} & 0.01 & 1\phantom{00} & 0.01 & 1\phantom{00} & 0.01\\
\hline
\end{tabular*}
\end{table}


\subsubsection*{Quality of partitions (Table \textup{\protect\ref{table:analyseM}})} We also
focus on the Rand Index for each algorithm. Indeed, even if poor
estimation of
$\lambda$ reveals a small Rand Index (Table~\ref{table:analyseM}),
good estimates do not always
lead to correctly estimated partitions. An illustration is given with
model 3 for which algorithms
produce good estimates with poor Rand Index, due to the nonmodular
structure of the network. As expected, the performance increases with
the number of nodes
(Table \ref{table:analyseV}).


\begin{table}[b]
\caption{Means and standard deviations of the
Rand Index with speed of the algorithms. $q=5$, model 2}\label{table:analyseV}
\begin{tabular*}{\textwidth}{@{\extracolsep{\fill}}ld{3.2}cd{3.2}cd{3.2}cd{5.2}d{2.2}@{}}
\hline
& \multicolumn{2}{c}{\textbf{Online-SAEM}} & \multicolumn{2}{c}{\textbf{Online-variational}} & \multicolumn{2}{c}{\textbf{Online-CEM}} & \multicolumn
{2}{c@{}}{\textbf{Batch-MixNet}} \\[-6pt]
& \multicolumn{2}{c}{\hrulefill} & \multicolumn{2}{c}{\hrulefill} & \multicolumn{2}{c}{\hrulefill} & \multicolumn
{2}{c@{}}{\hrulefill} \\
\multicolumn{1}{@{}l}{$\bolds{n}$} & \multicolumn{1}{c}{$\bolds{\overline{\mathit{rand}}}$} &
\multicolumn{1}{c}{$\bolds{\sigma_{\mathit{rand}}}$} &
\multicolumn{1}{c}{$\bolds{\overline{\mathit{rand}}}$} &
\multicolumn{1}{c}{$\bolds{\sigma_{\mathit{rand}}}$} &
\multicolumn{1}{c}{$\bolds{\overline{\mathit{rand}}}$} &
\multicolumn{1}{c}{$\bolds{\sigma_{\mathit{rand}}}$} &
\multicolumn{1}{c}{$\bolds{\overline{\mathit{rand}}}$} &
\multicolumn{1}{c@{}}{$\bolds{\sigma_{\mathit{rand}}}$} \\
\hline
\phantom{0}100 & 0.15 & 0.04 & 0.15 & 0.07 & 0.15 & 0.05 & 0.19 & 0.09 \\
\phantom{0}250 & 0.50 & 0.09 & 0.55 & 0.11 & 0.51 & 0.01 & 0.95 & 0.07 \\
\phantom{0}500 & 0.62 & 0.09 & 0.62 & 0.11 & 0.65 & 0.14 & 1 & 0.00 \\
\phantom{0}750 & 0.84 & 0.03 & 0.85 & 0.03 & 0.84 & 0.04 & 1 & 0.00 \\
1000 & 0.94 & 0.01 & 0.95 & 0.01 & 0.92 & 9.37 & 1 & 0.00 \\
2000 & 0.98 & 0.00 & 0.98 & 0.01 & 0.98 & 0.01 & 1 & 0.00 \\[12pt]
&
\multicolumn{1}{c}{$\overline{\mathit{time}}$} &
\multicolumn{1}{c}{$\sigma_{\mathit{time}}$} &
\multicolumn{1}{c}{$\overline{\mathit{time}}$} &
\multicolumn{1}{c}{$\sigma_{\mathit{time}}$} &
\multicolumn{1}{c}{$\overline{\mathit{time}}$} &
\multicolumn{1}{c}{$\sigma_{\mathit{time}}$} &
\multicolumn{1}{c}{$\overline{\mathit{time}}$} &
\multicolumn{1}{c@{}}{$\sigma_{\mathit{time}}$} \\
\phantom{0}100 & 0.09 & 0.00 & 0.09 & 0.00 & 0.09 & 0.00 & 0.10 & 0.00 \\
\phantom{0}250 & 1.31 & 0.01 & 1.32 & 0.01 & 1.31 & 0.00 & 3.18 & 0.01 \\
\phantom{0}500 & 1.41 & 0.01 & 1.46 & 0.01 & 1.41 & 0.01 & 49.46 & 0.13 \\
\phantom{0}750 & 3.45 & 0.02 & 3.57 & 0.02 & 3.44 & 0.02 & 251.32 & 0.75 \\
1000 & 9.46 & 0.41 & 9.61 & 0.43 & 9.37 & 0.40 & 805.92 & 0.49 \\
2000 & 157.31 & 1.28 & 158.21& 1.41 & 157.12 & 2.08 & 13051.10 & 73.75
\\
\hline
\end{tabular*}
\end{table}

\subsubsection*{Computational efficiency (Table \textup{\protect\ref{table:analyseV}})}
Since the aim of online methods is to provide computationally efficient
algorithms, the performance mentioned above should be put in
perspective with the speed of execution of each algorithm. Indeed,
Table \ref{table:analyseV} shows the strong gain of speed provided by
online methods compared with the batch algorithm. The speed of
execution is divided by 100 on networks with 2000 nodes, for instance.
Table \ref{table:analyseV} also shows that there is no
significant difference in the speed of execution among online methods.
Since the online variational method provides the best results in terms
of estimation precision, with no significant
difference with other methods on partition quality or speed, this will
be the algorithm chosen for the following.

\subsubsection*{Comparison with other algorithms (Table \textup{\protect\ref{table:res2}})}
The above results show that a strong case may be made for the online
variational algorithm when choosing
between alternative clustering methods. Consequently, we shall now
compare it with two suitable ``rivals'' for large networks:
a basic spectral clustering algorithm [\citet{Ng2002}], and one of the
popular community detection algorithms [\citet{Newman2006}].
The spectral clustering algorithm searches for a partition in the
space spanned by the eigenvectors
of the normalized Laplacian, whereas the community detection algorithm
looks for modules which are defined
by high intra-connectivity and low inter-connectivity.

For our five models with arbitrary fixed parameters $n=1000$, $Q=3$,
we ran these algorithms and computed the Rand Index
for each of them. From Table \ref{table:res2} we see that our online
variational algorithm always produces the
best clustering of nodes.

\begin{table}
\caption{Means and standard deviation of the Rand
Index for the five models computed over 30 different
runs for graph clustering competitors and variational algorithms}\label{table:res2}
\begin{tabular*}{\textwidth}{@{\extracolsep{4in minus 4in}}lcccccc@{}}
\hline
& \multicolumn{2}{c@{}}{\textbf{Community detection}} & \multicolumn
{2}{c}{\textbf{Spectral clustering}} & \multicolumn{2}{c@{}}{\textbf{Online-variational}}
\\[-6pt]
 & \multicolumn{2}{c@{}}{\rule{82pt}{0.5pt}} & \multicolumn
{2}{c}{\hrulefill} & \multicolumn{2}{c@{}}{\hrulefill}
\\
\textbf{Model}  & \multicolumn{1}{c}{$\bolds{\overline{\mathit{rand}}}$} &
 \multicolumn{1}{c}{$\bolds{\sigma_{\mathit{rand}}}$} &
 \multicolumn{1}{c}{$\bolds{\overline{\mathit{rand}}}$} &
\multicolumn{1}{c}{$\bolds{\sigma_{\mathit{rand}}}$} &
\multicolumn{1}{c}{$\bolds{\overline{\mathit{rand}}}$} &
\multicolumn{1}{c@{}}{$\bolds{\sigma_{\mathit{rand}}}$}\\
\hline
1 & 1.00 & 0.00 & 0.97 & 0.14 & 1.00 & 0.00 \\
2 & 0.99 & 0.01 & 0.98 & 0.00 & 1.00 & 0.00 \\
3 & 0.97 & 0.02 & 0.97 & 0.00 & 1.00 & 0.00 \\
4 & 0.00 & 0.00 & 0.00 & 0.00 & 0.00 & 0.00 \\
5 & 0.00 & 0.00 & 0.92 & 0.19 & 1.00 & 0.00 \\
\hline
\end{tabular*}
\end{table}
We generated networks using the MixNet data generating process. Thus,
these results correspond to what may be expected on networks that display
a blockmodel structure: the online variational algorithm always yields
the best node classification. Apart from model 4, it will also be
remarked that the spectral algorithm is fairly efficient
with a slight bias, and so the spectral clustering algorithm is
consistently more accurate than the community algorithm, the latter
failing completely when applied to model 5. Although the community
algorithm appears less well adapted to these experiments, we shall
see in the next section that this algorithm is particularly suitable
when partitioning data sets whose nodes are densely interconnected.

\subsection{Realistic networks growing over time}

\begin{figure}

\includegraphics{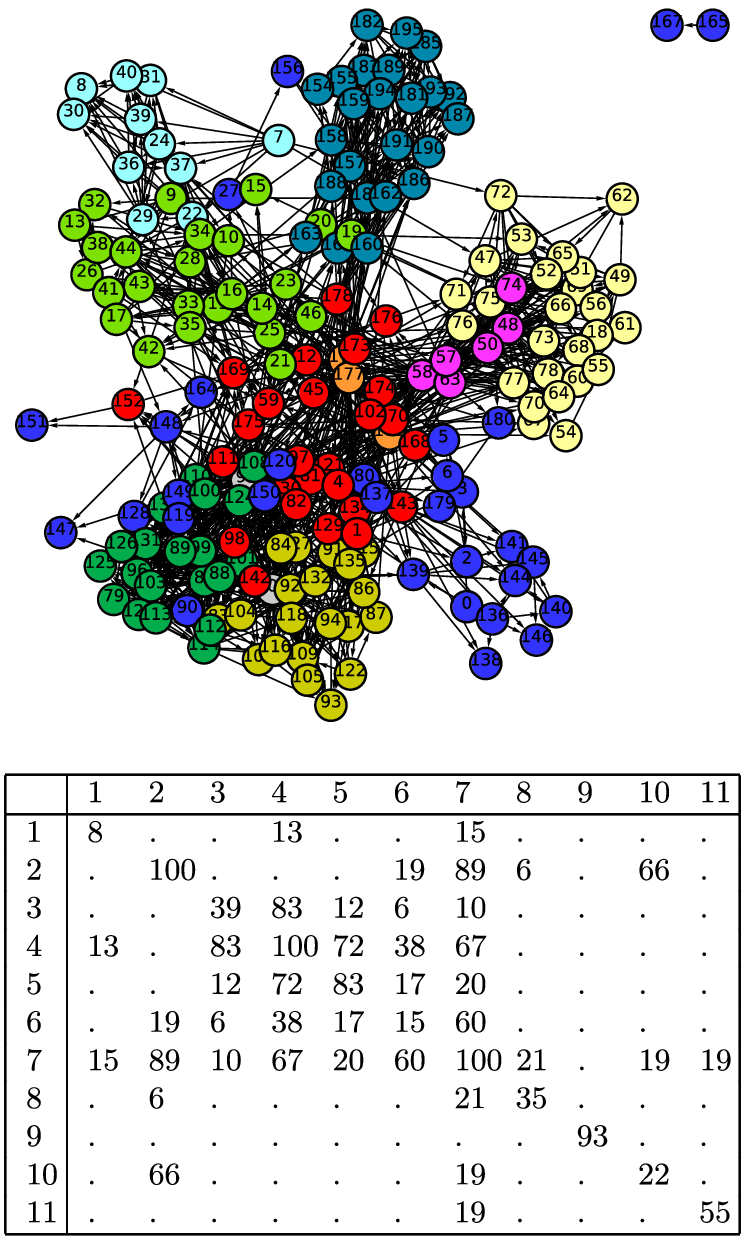}

\caption{MixNet results display on the French
political Blogosphere represented with the
organic layout of Cytoscape \protect\citep{SMO03}. The table corresponds to
the probabilities ($\times\!100$) of connection between the 11 selected
clusters [using a penalized likelihood criterion described in \protect\citet
{Daudin2008}]. Dots in the table correspond to connections lower than
$1\%$.} \label{fig:frenchblog}
\end{figure}

In this section we use a real network as a template to simulate a
realistic complex structure.
For this purpose, we use a French Political Blogosphere network data
set that consists of a sample of 196 political blogs from
a single day snapshot. This network was automatically extracted
October 14, 2006 and manually classified by the
``Observatoire Presidentielle'' project. This project is the result of
a collaboration between
RTGI SAS and Exalead and aims at analyzing the French presidential
campaign on the web. In this data set,
nodes represent hostnames (a hostname contains a set of pages) and
edges represent hyperlinks between different hostnames. If several
links exist between two different
hostnames, we collapse them into a single one. Note that intra-domain
links can be considered if
hostnames are not identical. Finally, in this experimentation we
consider that edges are not
oriented, which is not realistic but which does not affect the
interpretation of the groups. Six known communities compose this
network: Gauche (French Democrat), Divers Centre (Moderate party),
Droite (French Republican), Ecologiste (Green), Liberal (supporters of
economic-liberalism) and, finally, Analysts. The data is provided within
the \texttt{MixeR} package. This network presents an interesting
organization due to the existence of several political parties and
commentators. This complex
connectivity pattern is enhanced by MixNet parameters given in Figure
\ref{fig:frenchblog}.

%


As the algorithm is motivated by large data sets, we use the
parameters given by MixNet to generate networks
that grow over time. We use this French Blog to generate a realistic
network structure as a start point. We simulate
200 nodes networks from this model, then we iterate by simulating the
growth over time of these networks according to
the same model and we use the online algorithm to update parameters
sequentially. The result is striking: even on very
large networks with $\sim$13,000 nodes and $\sim$13,000,000 edges,
the online algorithm allows us to estimate mixture parameters
with negligible classification error in $\sim$6 minutes (Table~\ref
{table:grow}). This is the only algorithmic framework that allows
to perform model based clustering on networks of that size.

\section{Application to the 2008 US Presidential WebSphere}

Since its creation and enhanced by its recent social aspect
(Web 2.0), the World Wide Web is the space where
individuals use Internet technologies to talk, discuss and debate.
Such space can be seen as a directed graph
where the pages and hyperlinks are respectively represented by nodes
and edges. From this graph, many studies,
like \citet{broder2000graph}, have been published and introduced the
key properties of the Web structure. However, this section rather
focuses on local studies by considering
that the Web is formed by \textit{territories} and \textit
{communities} with their own conversation leaders and
participants [\citet{ghitalla-outre}]. Here, we define a territory as a
group of websites concerned by the
same topic and a community as a group of websites in the same
territory which may share the same opinion
or the same link connectivity. One usually assumes that the existence
of a hyperlink between two pages
implies that they are content-related [\citet
{kleinberg1998}; \citet{davison2000}]. By exploring the link page exchanges, one
can actually draw the borders of
web territories/communities.

\subsubsection*{Comparison with a community detection algorithm} A first
step consists in comparing the results of MixNet
with the community detection algorithm proposed by \citet{Newman2006}.
If the political classification is used as a
reference, the community algorithm produces better agreement with a
$\mathit{randIndex} = 0.59$, compared with a $\mathit{randIndex}=0.25$ for MixNet (see Table \ref{table:Contingency3}). However,
it appears that this comparison favors Newman, whereas the methods
have different objective. Indeed, the community algorithm aims at
finding modules which are defined by high intra-connectivity and low
inter-connectivity. Given that websites tend to link to one another
in line with political affinities, the link topology corresponding to
the manual classification naturally favors the
community module definition. The objective function can also help to
explain the community algorithm's suitability for this data set, since
the quality of a
partition in terms of Newman's modules can be expressed in terms of
the \textit{modularity}, which is maximized. The value of this
modularity is a
scalar between $-$1 and 1 and measures the density of links inside
communities as compared to links between communities [\citet{Newman2006}].
When applying both algorithms on our political network with $Q=3$, the
online variational algorithm yields a $\mathit{modularity} = 0.20$, whereas
the community algorithm yields a $\mathit{modularity} = 0.30$, which is close
to the manual partition modularity of 0.28. As MixNet classes do not
necessarily take the form of modules, one might expect our approach to
yield a modularity index that is not ``optimal.'' Nevertheless, the two
class definitions are complementary, and both are needed in order to
give a global overview of a network: the \textit{community} partition to detect
dense node connectivity, and the MixNet partition to analyze nodes
with similar connectivity profiles. However, as mentioned by
\citet{adamic2005pba}, the division between liberal and conservative
blogs is ``unmistakable,'' this is why it may be more interesting to
uncover the structure of the two communities rather than detecting them.

\begin{table}
\caption{Quality of the clustering procedure in
terms of Rand Index when the network grows over time. Each
configuration has been simulated 100 times}\label{table:grow}
\begin{tabular*}{\textwidth}{@{\extracolsep{\fill}}lcd{1.3}d{3.1}@{}}
\hline
 \multicolumn{1}{@{}l}{$\bolds{\#}$ \textbf{nodes (previous${}\bolds{+}{}$new)}} &  \multicolumn{1}{c}{\phantom{00.}\textbf{Ave.} $\bolds{\#}$ \textbf{edges}} &
  \multicolumn{1}{c}{\textbf{Ave. rand}} &  \multicolumn{1}{c@{}}{\textbf{Ave. cpu
time (s)}}\\
\hline
\phantom{0}200 & \phantom{00000.00}3131.72 & 0.94 & 0.9\\
\phantom{0}200 $+$ 200 & \phantom{000000}50,316.32 & 0.998 & 0.4\\
\phantom{0}400 $+$ 400 & \phantom{000000}12,486.24 & 0.999 & 1.4\\
\phantom{0}800 $+$ 800 & \phantom{0000}201,009.5 & 1 & 5.7\\
1600 $+$ 1600 & \phantom{0000}803,179.6 & 1 & 22.8\\
3200 $+$ 3200 & \phantom{0}3,202,196 & 1 & 91.9\\
6400 $+$ 6400 & 12,804,008 & 1 & 371.1\\
\hline
\end{tabular*}
\end{table}
%

\begin{table}[b]
\caption{Contingency table comparing the
political partition and MixNet partition}\label{table:Contingency3}
\begin{tabular*}{200pt}{@{\extracolsep{\fill}}lccc@{}}
\hline
& \textbf{Conservative} & \textbf{Independent} & \textbf{Liberal} \\
\hline
Cluster 1 & 734 & 135 & 238 \\
Cluster 2 & 290 & \phantom{0}26 & \phantom{00}8 \\
Cluster 3 & \phantom{00}2 & \phantom{00}7 & 430 \\
\hline
\end{tabular*}

\end{table}

\subsubsection*{Interpreting MixNet results} 
\begin{figure}

\includegraphics{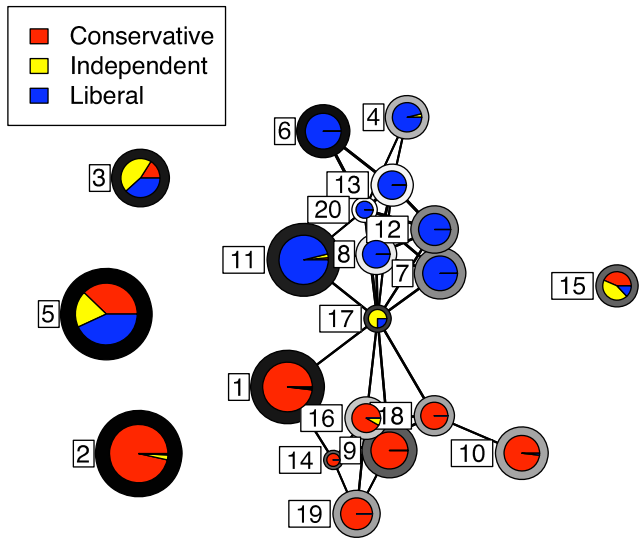}

\caption{Network summary of US political
websites. Each
vertex represents a cluster. Each pie chart gives the proportions of
liberal, conservative and independent tagged websites in the cluster.
The outer
ring color of the vertices is proportional to the intensity of the
intra-connectivity: the darker, the weaker. Edges are
represented when the inter-connectivity is among the 20\% of the largest
among all connectivity values.} \label{fig:=graphUS}
\end{figure}

MixNet first confirms what was
already mentioned by \citet{adamic2005pba}:
the political websphere is partioned according to political
orientations. In addition, MixNet highlights the role of main US online
portals as the core of this websphere (Figure \ref{fig:=graphUS},
C17). Political communities do not directly cite
their opponents but communicate through \href{http://www.nytimes.com/}{nytimes.com},
\href{http://www.washingtonpost.com/}{washingtonpost.com}, \href{http://edition.cnn.com/}{cnn.com} or \href{http://www.msn.com/}{msn.com},
for instance (in C17). This central structure has two main
significations: it confirms the political
cyberbalkanization trend that was already observed in 2004, and it
emphasizes the role of mass media websites as
political referees. Plus, the connectivity pattern estimated by the
model shows a particular affinity between the mass-media
cluster with the liberal thought, as connections are stronger toward
the liberal part of the weblogs (Table~\ref{table:PI}).

\begin{table}
\tabcolsep=0pt
\caption{Estimated $\pi$ (in percentage) and number of nodes in each
cluster for the US political websphere. $\bar{d}$~represents the
estimated mean degree of each group. Clusters with probabilities of
connection lower than $1\%$ are not represented (clusters 3, 5, 15)}\label{table:PI}
\begin{tabular*}{\textwidth}{@{\extracolsep{\fill}}ld{3.0}d{3.0}d{3.0}d{3.0}d{3.0}d{3.0}d{3.0}d{3.0}d{3.0}d{3.0}d{3.0}d{3.0}d{3.0}d{3.0}d{3.0}d{3.0}d{3.0}@{}}
\hline
 & \multicolumn{8}{c}{\hspace*{40pt}\textbf{Conservative}} & \multicolumn{8}{c@{}}{\hspace*{40pt}\textbf{Liberal}}
\\[-6pt]
 & &\multicolumn{8}{c}{\hrulefill} & \multicolumn{8}{c@{}}{\hrulefill} \\
\textbf{ID} & \textbf{17} & \textbf{1} & \textbf{2} & \textbf{10} & \textbf{9} & \textbf{14} & \textbf{16} & \textbf{18} & \textbf{19} &
 \textbf{11} & \textbf{4} & \textbf{6} & \textbf{7} & \textbf{8} & \textbf{12}
& \textbf{13} & \multicolumn{1}{c@{}}{\textbf{20}} \\
\hline
17 &\multicolumn{1}{c}{$\cdot$} & 54 &\multicolumn{1}{c}{$\cdot$} & \multicolumn{1}{c}{$\cdot$} & 64 &\multicolumn{1}{c}{$\cdot$} & 52 & 66 &\multicolumn{1}{c}{$\cdot$} & 62 &\multicolumn{1}{c}{$\cdot$} &\multicolumn{1}{c}{$\cdot$} & 67 & 67 & 62
& 43 & 100 \\[3pt]
\phantom{0}1 & 54 &\multicolumn{1}{c}{$\cdot$} &\multicolumn{1}{c}{$\cdot$} &\multicolumn{1}{c}{$\cdot$} &\multicolumn{1}{c}{$\cdot$} & 54 &\multicolumn{1}{c}{$\cdot$} &\multicolumn{1}{c}{$\cdot$} &\multicolumn{1}{c}{$\cdot$} &\multicolumn{1}{c}{$\cdot$} &\multicolumn{1}{c}{$\cdot$} &\multicolumn{1}{c}{$\cdot$} &\multicolumn{1}{c}{$\cdot$} &\multicolumn{1}{c}{$\cdot$} &\multicolumn{1}{c}{$\cdot$} &\multicolumn{1}{c}{$\cdot$} &
\multicolumn{1}{c}{$\cdot$} \\
\phantom{0}2 &\multicolumn{1}{c}{$\cdot$} &\multicolumn{1}{c}{$\cdot$} &\multicolumn{1}{c}{$\cdot$} &\multicolumn{1}{c}{$\cdot$} &\multicolumn{1}{c}{$\cdot$} &\multicolumn{1}{c}{$\cdot$} &\multicolumn{1}{c}{$\cdot$} &\multicolumn{1}{c}{$\cdot$} &\multicolumn{1}{c}{$\cdot$} &\multicolumn{1}{c}{$\cdot$} &\multicolumn{1}{c}{$\cdot$} &\multicolumn{1}{c}{$\cdot$} &\multicolumn{1}{c}{$\cdot$} &\multicolumn{1}{c}{$\cdot$} &\multicolumn{1}{c}{$\cdot$} &\multicolumn{1}{c}{$\cdot$} &\multicolumn{1}{c}{$\cdot$}
\\
10 &\multicolumn{1}{c}{$\cdot$} &\multicolumn{1}{c}{$\cdot$} &\multicolumn{1}{c}{$\cdot$} & 58 &\multicolumn{1}{c}{$\cdot$} &\multicolumn{1}{c}{$\cdot$} &\multicolumn{1}{c}{$\cdot$} & 47 &\multicolumn{1}{c}{$\cdot$} &\multicolumn{1}{c}{$\cdot$} &\multicolumn{1}{c}{$\cdot$} &\multicolumn{1}{c}{$\cdot$} &\multicolumn{1}{c}{$\cdot$} &\multicolumn{1}{c}{$\cdot$} &\multicolumn{1}{c}{$\cdot$} &\multicolumn{1}{c}{$\cdot$} &
\multicolumn{1}{c}{$\cdot$} \\
\phantom{0}9 & 64 &\multicolumn{1}{c}{$\cdot$} &\multicolumn{1}{c}{$\cdot$} &\multicolumn{1}{c}{$\cdot$} &\multicolumn{1}{c}{$\cdot$} & 56 & 61 &\multicolumn{1}{c}{$\cdot$} & 40 &\multicolumn{1}{c}{$\cdot$} &\multicolumn{1}{c}{$\cdot$} &\multicolumn{1}{c}{$\cdot$} &\multicolumn{1}{c}{$\cdot$} &\multicolumn{1}{c}{$\cdot$} &\multicolumn{1}{c}{$\cdot$} &\multicolumn{1}{c}{$\cdot$}
&\multicolumn{1}{c}{$\cdot$} \\
14 &\multicolumn{1}{c}{$\cdot$} & 54 &\multicolumn{1}{c}{$\cdot$} &\multicolumn{1}{c}{$\cdot$} & 56 &\multicolumn{1}{c}{$\cdot$} & 55 & 72 & 40 &\multicolumn{1}{c}{$\cdot$} &\multicolumn{1}{c}{$\cdot$} &\multicolumn{1}{c}{$\cdot$} &\multicolumn{1}{c}{$\cdot$} &\multicolumn{1}{c}{$\cdot$} &\multicolumn{1}{c}{$\cdot$} &
\multicolumn{1}{c}{$\cdot$} &\multicolumn{1}{c}{$\cdot$} \\
16 & 52 &\multicolumn{1}{c}{$\cdot$} &\multicolumn{1}{c}{$\cdot$} &\multicolumn{1}{c}{$\cdot$} & 61 & 55 & 73 & 60 & 56 &\multicolumn{1}{c}{$\cdot$} &\multicolumn{1}{c}{$\cdot$} &\multicolumn{1}{c}{$\cdot$} &\multicolumn{1}{c}{$\cdot$} &\multicolumn{1}{c}{$\cdot$} &\multicolumn{1}{c}{$\cdot$} &
\multicolumn{1}{c}{$\cdot$} &\multicolumn{1}{c}{$\cdot$} \\
18 & 66 &\multicolumn{1}{c}{$\cdot$} &\multicolumn{1}{c}{$\cdot$} & 47 &\multicolumn{1}{c}{$\cdot$} & 72 & 60 & 58 &\multicolumn{1}{c}{$\cdot$} &\multicolumn{1}{c}{$\cdot$} &\multicolumn{1}{c}{$\cdot$} &\multicolumn{1}{c}{$\cdot$} &\multicolumn{1}{c}{$\cdot$} &\multicolumn{1}{c}{$\cdot$} &\multicolumn{1}{c}{$\cdot$} &
\multicolumn{1}{c}{$\cdot$} &\multicolumn{1}{c}{$\cdot$} \\
19 &\multicolumn{1}{c}{$\cdot$} &\multicolumn{1}{c}{$\cdot$} &\multicolumn{1}{c}{$\cdot$} &\multicolumn{1}{c}{$\cdot$} & 40 & 40 & 56 &\multicolumn{1}{c}{$\cdot$} & 57 &\multicolumn{1}{c}{$\cdot$} &\multicolumn{1}{c}{$\cdot$} &\multicolumn{1}{c}{$\cdot$} &\multicolumn{1}{c}{$\cdot$} &\multicolumn{1}{c}{$\cdot$} &\multicolumn{1}{c}{$\cdot$} &\multicolumn{1}{c}{$\cdot$}
&\multicolumn{1}{c}{$\cdot$} \\[3pt]
11 & 62 &\multicolumn{1}{c}{$\cdot$} &\multicolumn{1}{c}{$\cdot$} &\multicolumn{1}{c}{$\cdot$} &\multicolumn{1}{c}{$\cdot$} &\multicolumn{1}{c}{$\cdot$} &\multicolumn{1}{c}{$\cdot$} &\multicolumn{1}{c}{$\cdot$} &\multicolumn{1}{c}{$\cdot$} &\multicolumn{1}{c}{$\cdot$} &\multicolumn{1}{c}{$\cdot$} &\multicolumn{1}{c}{$\cdot$} &\multicolumn{1}{c}{$\cdot$} & 47 &\multicolumn{1}{c}{$\cdot$} &\multicolumn{1}{c}{$\cdot$} &
40 \\
\phantom{0}4 &\multicolumn{1}{c}{$\cdot$} &\multicolumn{1}{c}{$\cdot$} &\multicolumn{1}{c}{$\cdot$} &\multicolumn{1}{c}{$\cdot$} &\multicolumn{1}{c}{$\cdot$} &\multicolumn{1}{c}{$\cdot$} &\multicolumn{1}{c}{$\cdot$} &\multicolumn{1}{c}{$\cdot$} &\multicolumn{1}{c}{$\cdot$} &\multicolumn{1}{c}{$\cdot$} & 65 &\multicolumn{1}{c}{$\cdot$} &\multicolumn{1}{c}{$\cdot$} &\multicolumn{1}{c}{$\cdot$} &\multicolumn{1}{c}{$\cdot$} & 42 &
40 \\
\phantom{0}6 &\multicolumn{1}{c}{$\cdot$} &\multicolumn{1}{c}{$\cdot$} &\multicolumn{1}{c}{$\cdot$} &\multicolumn{1}{c}{$\cdot$} &\multicolumn{1}{c}{$\cdot$} &\multicolumn{1}{c}{$\cdot$} &\multicolumn{1}{c}{$\cdot$} &\multicolumn{1}{c}{$\cdot$} &\multicolumn{1}{c}{$\cdot$} &\multicolumn{1}{c}{$\cdot$} &\multicolumn{1}{c}{$\cdot$} &\multicolumn{1}{c}{$\cdot$} &\multicolumn{1}{c}{$\cdot$} &\multicolumn{1}{c}{$\cdot$} &\multicolumn{1}{c}{$\cdot$} & 88 &
99 \\
\phantom{0}7 & 67 &\multicolumn{1}{c}{$\cdot$} &\multicolumn{1}{c}{$\cdot$} &\multicolumn{1}{c}{$\cdot$} &\multicolumn{1}{c}{$\cdot$} &\multicolumn{1}{c}{$\cdot$} &\multicolumn{1}{c}{$\cdot$} &\multicolumn{1}{c}{$\cdot$} &\multicolumn{1}{c}{$\cdot$} &\multicolumn{1}{c}{$\cdot$} &\multicolumn{1}{c}{$\cdot$} &\multicolumn{1}{c}{$\cdot$} & 47 & 76 & 49 &\multicolumn{1}{c}{$\cdot$}
& 76 \\
\phantom{0}8 & 67 &\multicolumn{1}{c}{$\cdot$} &\multicolumn{1}{c}{$\cdot$} &\multicolumn{1}{c}{$\cdot$} &\multicolumn{1}{c}{$\cdot$} &\multicolumn{1}{c}{$\cdot$} &\multicolumn{1}{c}{$\cdot$} &\multicolumn{1}{c}{$\cdot$} &\multicolumn{1}{c}{$\cdot$} & 47 &\multicolumn{1}{c}{$\cdot$} &\multicolumn{1}{c}{$\cdot$} & 76 & 90 & 74 &
81 & 98 \\
12 & 62 &\multicolumn{1}{c}{$\cdot$} &\multicolumn{1}{c}{$\cdot$} &\multicolumn{1}{c}{$\cdot$} &\multicolumn{1}{c}{$\cdot$} &\multicolumn{1}{c}{$\cdot$} &\multicolumn{1}{c}{$\cdot$} &\multicolumn{1}{c}{$\cdot$} &\multicolumn{1}{c}{$\cdot$} &\multicolumn{1}{c}{$\cdot$} &\multicolumn{1}{c}{$\cdot$} &\multicolumn{1}{c}{$\cdot$} & 49 & 74 & 45 &
92 & 98 \\
13 & 43 &\multicolumn{1}{c}{$\cdot$} &\multicolumn{1}{c}{$\cdot$} &\multicolumn{1}{c}{$\cdot$} &\multicolumn{1}{c}{$\cdot$} &\multicolumn{1}{c}{$\cdot$} &\multicolumn{1}{c}{$\cdot$} &\multicolumn{1}{c}{$\cdot$} &\multicolumn{1}{c}{$\cdot$} &\multicolumn{1}{c}{$\cdot$} & 42 & 88 &\multicolumn{1}{c}{$\cdot$} & 81 & 92 &
95 & 99 \\
20 & 100 &\multicolumn{1}{c}{$\cdot$} &\multicolumn{1}{c}{$\cdot$} &\multicolumn{1}{c}{$\cdot$} &\multicolumn{1}{c}{$\cdot$} &\multicolumn{1}{c}{$\cdot$} &\multicolumn{1}{c}{$\cdot$} &\multicolumn{1}{c}{$\cdot$} &\multicolumn{1}{c}{$\cdot$} & 40 & 42 & 99 & 76 & 98 & 98
& 99 & 100 \\[6pt]
$N_q$ & 4 & 214 & 407 & 66 & 56 & 1 & 24 & 19 & 36 & 26 & 58 & 207 &
51 & 20 & 37 & 23 & 3 \\
$\bar{d}$ & 649 & 86 & 28 & 149 & 69 & 455 & 335 & 167 & 172 & 192 &
64 & 66 & 66 & 310& 154 & 170 & 324 \\
\hline
\end{tabular*}
\end{table}

Then the question is to determine what are the structural
characteristics of the liberal and conservative
territories (note that independent sites do not seem to be structured
on their own). MixNet reveals a hierarchical
organization of political sub-spheres with weblogs having a
determinant role in the structuration of
the liberal community, \href{http://reachm.com/}{reachm.com}, \href{http://www.mahablog.com/}{mahablog.com},
\href{http://www.juancole.com/}{juancole.com} (C20), which are well known to be at
the core of the liberal debate on the web. This results in a set of
clusters (C7, C8, C12, C13 and C20) that
show very strong intra $and$ inter group connectivities which nearly
forms a clique (Table \ref{table:PI}). The
balkanization is also observed within territories, as radical
positions, like in the \href{http://feministe.us/}{feministe.us} website~(C6), are only spread through core websites ($\hat{\pi}_{20,6}=99\%$,
for instance). A last level of hierarchy is
made by liberal blogs that show intermediate connections within the
same liberal territory.

Interestingly, this subdivision is also present in the conservative
part of the network, with very famous websites like
\href{http://www.foxnews.com/}{foxnews.com} (C14) being at the center of the debate. Indeed,
clusters C3, C14, C16, C18 and C19 constitute
the core of the conservative websphere, and clusters C1 and C2 are
very lightly connected with other conservative blogs.
The difference lies in the intensity of connection, which is lower for
the conservatives.

\begin{figure}

\includegraphics{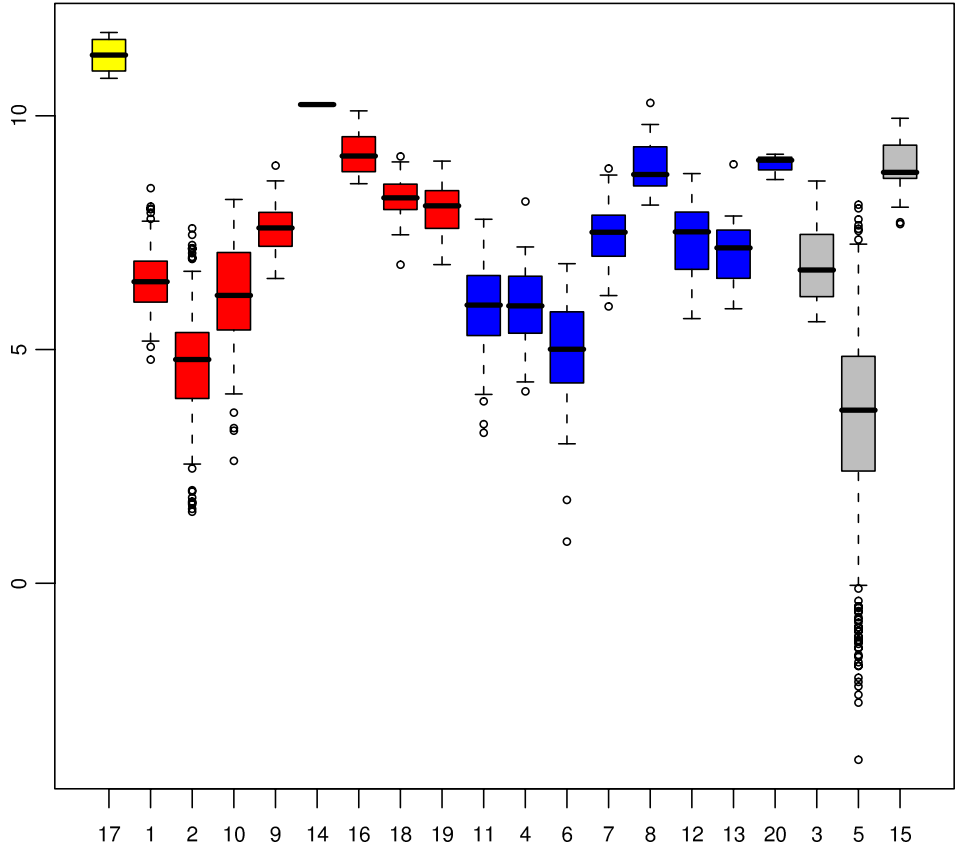}

\caption{Boxplot of MixNet classes betweenness (in log).}\label{fig:between}
\end{figure}

Compared with available methods that can analyze networks of such size
(like community detection), MixNet shows structures of the
political websphere that are more complex than the expected
liberal/conservative split. The model highlights the structural
similarities that exist between spheres of political opponents. Both
communities are characterized by a small
set of sites which use the internet in a very professional and
efficient way, with a lot of cross-linking. This results
in a core structure to which other sites are linked, these other sites
being less efficient in the citations to other
websites. This could be explained either by a tendency to ignore other
elements in the debate or by a use of the internet
which is less efficient. Interestingly, this structure is very similar
between conservatives and liberals, with the liberal
core being more tight. For the liberal blogs, this observation can
result from a better understanding of
their Web Ecosystem. This interpretation is reinforced by the
different betweenness centralities of MixNet classes.
Betweenness is based on the number of shortest geodesic paths that
pass through a vertex. Figure~\ref{fig:between}
shows that MixNet betweenness is higher for MixNet core classes on
average in both political structures, whereas
the betweenness patterns of the liberals and conservatives look very similar.

\section{Conclusion}

In this paper we propose an online version of estimation algorithms for
random graphs which are based on a mixture of distributions. These
strategies allow the estimation of model parameters within a reasonable
computation time for data sets which can be made up of thousands of
nodes. These methods constitute a trade-off between the potential
amount of data to process and the quality of the estimations: even if
online methods are not as precise as batch methods for estimation,
they may represent a solution when the size of the network is too
large for any existing estimation strategy. Furthermore, our
simulation study shows that the quality of the remaining partition is
good when using online methods. In the network of 2008 US political
websites, we could uncover the structure that makes the political
websphere. This structure is very different from classical modules
or ``communities,'' which highlights the need for efficient
computational strategies
to perform model-based clustering on large graphs. The online
framework is very flexible,
and could be applied to other models such as the block model and the
mixed membership model,
as the online framework can be adapted to Bayesian algorithms [\citet{Opper}].


\begin{appendix}
\section*{Appendix}\label{app}

\subsection{Examples of distributions for the exponential family}

We provide some examples of common distributions that can be used in
the context of networks.
For example, when the only available information is the presence or
the absence of an edge, then $X_{ij}$ is assumed to follow
a Bernoulli distribution:
\[
X_{ij} | Z_{iq}Z_{jl}=1 \sim\mathcal{B}(\pi_{ql})
\cases{
\eta_{ql} = \log\displaystyle\frac{\pi_{ql}}{1-\pi_{ql}}, \vspace*{2pt}\cr
h(X_{ij})=X_{ij},\vspace*{2pt}\cr
a(\eta_{ql}) = \log(1-\pi_{ql}),\vspace*{2pt}\cr
b(X_{ij})=0.}
\]
If additional information is available to describe the connections
between vertices, it may be integrated into the model. For example,
the Poisson distribution might describe the intensity of the traffic
between nodes. A typical example in web access log mining is
the number of users going from a page $i$ to a page $j$. Another
example is provided by co-authorship networks, for which valuation
may describe the number of articles commonly published by the authors
of the network. In those cases, we have
\[
X_{ij} | Z_{iq}Z_{jl}=1 \sim\mathcal{P}(\lambda_{ql})
\cases{
\eta_{ql} = \log\lambda_{ql}, \vspace*{2pt}\cr
h(X_{ij})=X_{ij},\vspace*{2pt}\cr
a(\eta_{ql}) = - \lambda_{ql},\vspace*{2pt}\cr
b(X_{ij})=X_{ij}!
}
\]

\subsection{Parameters update in the Bernoulli and Poisson cases for
the online SAEM}
The estimator becomes
\[
\pi_{ql}^{[n+1]} = \gamma_{ql}^{[n+1]} \pi_{ql}^{[n] } + \bigl(1-\gamma
_{ql}^{[n+1]}\bigr) \frac{\xi_{ql}^{[n+1]} }{\zeta_{ql}^{[n+1]}},
\]
where
\begin{eqnarray*}
\gamma_{ql}^{[n+1]} & = & \frac{ N_q(\Zbf^{[n]}) N_l(\Zbf^{[n]})}{
N_q(\Zbf^{[n]}) N_l(\Zbf^{[n]}) + Z_{n+1,q} N_l(\Zbf^{[n]}) + Z_{n+1,l}
N_q(\Zbf^{[n]})},\\
\xi_{ql}^{[n+1]} & = & Z_{n+1,q} \sum_{j=1}^{n} Z_{jl}^{[n]} X_{n+1,j}
+ Z_{n+1,l} \sum_{i=1}^n Z_{iq}^{[n]} X_{i,n+1}, \\
\zeta_{ql}^{[n+1]} & = & Z_{n+1,q} N_l\bigl(\Zbf^{[n]}\bigr) + Z_{n+1,l} N_q\bigl(\Zbf
^{[n]}\bigr) + Z_{n+1,q} \mathbb{I}\{q=l\}.
\end{eqnarray*}

\subsection{Parameters update in the Bernoulli and the Poisson cases
for the online variational algorithm}
We get the following update equation:
\[
\pi_{ql}^{[n+1]} = \gamma_{ql}^{[n+1]} \pi_{ql}^{[n] } + \bigl(1-\gamma
_{ql}^{[n+1]}\bigr) \frac{ \Esp_{\mathcal{R}^{[n]}} ( \xi_{ql}^{[n+1]}
)}{\Esp_{\mathcal{R}^{[n]}} ( \zeta_{ql}^{[n+1]})},
\]
where
\begin{eqnarray*}
\gamma_{ql}^{[n+1]} & = & { \Esp_{\mathcal{R}^{[n]}}
\bigl(N_q\bigl(\Zbf^{[n]}\bigr)\bigr) \Esp_{\mathcal{R}^{[n]}} \bigl(N_l\bigl(\Zbf
^{[n]}\bigr)\bigr)}\\
&&{}/ \Esp_{\mathcal{R}^{[n]}} \bigl(N_q\bigl(\Zbf^{[n]}\bigr)\bigr)
\Esp_{\mathcal{R}^{[n]}} \bigl(N_l\bigl(\Zbf^{[n]}\bigr)\bigr) + \tau_{n+1,q}
\Esp_{\mathcal{R}^{[n]}} \bigl(N_l\bigl(\Zbf^{[n]}\bigr)\bigr) \\
&&{}+ \tau_{n+1,l}
\Esp_{\mathcal{R}^{[n]}} \bigl(N_q\bigl(\Zbf^{[n]}\bigr)\bigr),\\
\Esp_{\mathcal{R}^{[n]}}\bigl ( \xi_{ql}^{[n+1]} \bigr) & = & \tau
_{n+1,q} \sum_{j=1}^{n} \tau_{jl}^{[n]} X_{n+1,j} + \tau_{n+1,l} \sum_{i=1}^n
\tau_{iq}^{[n]} X_{i,n+1},\\
\Esp_{\mathcal{R}^{[n]}} \bigl( \zeta_{ql}^{[n+1]} \bigr) &=& \tau
_{n+1,q} \Esp_{\mathcal{R}^{[n]}} \bigl( N_l\bigl(\Zbf^{[n]}\bigr)\bigr) + \tau_{n+1,l}
\Esp_{\mathcal{R}^{[n]}} \bigl(N_q\bigl(\Zbf^{[n]}\bigr) \bigr)+ \tau_{n+1,q}
\mathbb{I}\{q=l\}
\end{eqnarray*}
with
\[
\Esp_{\mathcal{R}^{[n]}} \bigl( N_q\bigl(\Zbf^{[n]}\bigr)\bigr) = \sum_{i=1}^n
\tau_{iq}^{[n]}.
\]
\end{appendix}

\printaddresses

\end{document}